\documentclass[twocolumn,aps,superscriptaddress,nofootinbib,floatfix,linenumbers]{revtex4}

\usepackage{amsfonts}
\usepackage{float}
\usepackage{placeins}
\usepackage{graphicx}
\usepackage[hidelinks]{hyperref}
\usepackage{color,amsmath,amssymb,bm}
\usepackage{tensor}

\usepackage[normalem]{ulem}  

\renewcommand{\sout}{\bgroup \color{black} \ULdepth=-.5ex \ULset}

\def\blfootnote{\xdef\@thefnmark{}\@footnotetext}

\newcommand{\beq}{\begin{equation}}
\newcommand{\eeq}{\end{equation}}
\newcommand{\bea}{\begin{eqnarray}}
\newcommand{\eea}{\end{eqnarray}}

\newcommand{\gtsim}{\raisebox{-4pt}{$\,\stackrel{\textstyle >}{\sim}\,$}}

\begin{document}

\title{Quasi particle model vs lattice QCD thermodynamics: extension to $N_f=2+1+1$ flavors and momentum dependent quark masses}

\author{Maria Lucia Sambataro}
\email{sambataro@lns.infn.it}
\affiliation{Department of Physics and Astronomy, University of Catania, Via S. Sofia 64, 1-95125 Catania, Italy}
\affiliation{Laboratori Nazionali del Sud, INFN-LNS, Via S. Sofia 62, I-95123 Catania, Italy}

\author{Vincenzo Greco}
\email{vincenzo.greco@dfa.unict.it}
\affiliation{Department of Physics and Astronomy, University of Catania, Via S. Sofia 64, 1-95125 Catania, Italy}
\affiliation{Laboratori Nazionali del Sud, INFN-LNS, Via S. Sofia 62, I-95123 Catania, Italy}

\author{Gabriele Parisi}
\email{gabriele.parisi@dfa.unict.it}
\affiliation{Department of Physics and Astronomy, University of Catania, Via S. Sofia 64, 1-95125 Catania, Italy}
\affiliation{Laboratori Nazionali del Sud, INFN-LNS, Via S. Sofia 62, I-95123 Catania, Italy}

\author{Salvatore Plumari}
\email{salvatore.plumari@dfa.unict.it}
\affiliation{Department of Physics and Astronomy, University of Catania, Via S. Sofia 64, 1-95125 Catania, Italy}
\affiliation{Laboratori Nazionali del Sud, INFN-LNS, Via S. Sofia 62, I-95123 Catania, Italy}

\date{\today}

\begin{abstract}
In the last decade a Quasi-Particle Model ($QPM$) has supplied the basis for the study of HQ production in ultra-relativistic
AA collisions, allowing for a phenomenological estimate of the HQ diffusion coefficient $D_s(T)$.
Taking advantage of the new lattice QCD results for the Equation of State (EoS)  with 2+1+1 dynamical flavors, we extend our $QPM$ approach from $N_f=2+1$ to $N_f=2+1+1$, in which the charm quark is included. Given an effective coupling $g(T)$ fixed by a fit to the lQCD energy density $\epsilon(T)$, we evaluate the impact of different temperature parametrizations of charm quark mass on EoS and susceptibilities $\chi_q(T)$ of light, $\chi_s(T)$ of strange and $\chi_c(T)$ of charm quarks, the last favouring a charm quark mass increasing toward $T_c$. We also explore the extension of the $QPM$ approach to a more realistic approach, that we label $QPM_p$, in which quark and gluon masses explicitly depend on their momentum converging to the current quark mass at high momenta, as expected from asymptotic free dynamics. The $QPM_p$ is seen to allow for a simultaneous quantitative description not only of the EoS but also of the quark susceptibilities ($\chi_q(T)$, $\chi_s(T)$), which instead are underestimated in the simple $QPM$ model. Furthermore, evaluating the spatial diffusion coefficient $2\pi T D_s(T)$ in the $QPM_p$, we find it is also significantly closer than $QPM$ to the recent lQCD data performed including dynamical fermions. 
Finally, in a 1+1D expanding system, we evaluate the $R_{AA}(p_T)$ in the $QPM$ and $QPM_p$, finding a significant reduction at low momenta for $QPM_p$ which could lead in a realistic scenario to a better agreement to experimental data.

\end{abstract}

\maketitle

\section{Introduction}

The fundamental theory of strong interaction, Quantum Chromodynamics (QCD), can be solved numerically on a discretized lattice in the regime of vanishing or small values of the baryonic chemical potential $\mu_B$ \cite{Borsanyi:2013bia, Soltz:2015ula}. The study of strong interaction with finite baryon chemical potential is a challenging task, due to the fermionic sign problem. At low baryonic densities, the QCD predicts that the transition from hadrons to quarks is expected to be a crossover, while at high baryonic densities it is expected to become a first order phase transition \cite{Asakawa:1989bq, Fischer:2012vc, Herbst:2010rf, Schaefer:2007pw, Critelli:2017oub, Stephanov:1998dy}. 
\newline
In general, the study of the phase of matter made of quark and gluons, called Quark-Gluon Plasma (QGP),  is one the main goals of experiments at the Relativistic Heavy Ion Collider (RHIC) in Brookhaven National Laboratory and at the Large Hadron Collider (LHC) by ALICE, ATLAS and CMS collaborations at CERN. Many efforts have been done to discuss QCD thermodynamics by means of perturbation theory at high temperatures, but the perturbative expansion fails to reproduce the lattice data, such as the interaction measure $\langle\Theta\rangle_\mu^{\mu}=\epsilon - 3P$. In order to overcome this problem, some approaches  supplement pQCD with Hard-Thermal Loop (HTL) calculations \cite{Braaten:1991gm, Blaizot:1993be, Andersen:1999fw, Andersen:2002ey, Blaizot:1999ap, Andersen:2010wu}.
\newline
Another successful way to account for non-perturbative dynamics is given by the Quasi-Particle Model ($QPM$) \cite{Levai:1997yx, Peshier:2002ww, Plumari:2011mk, Song:2015ykw, Liu:2023rfi, Soloveva:2023tvj}, further extended including the off-shell dynamics, by the
Dynamic-QPM (DQPM) \cite{Peshier:2005pp, Cassing:2007nb, Berrehrah:2013mua,Berrehrah:2014kba}.
In such an approach, the strong interaction in a non-perturbative regime is considered through an effective temperature-dependent mass for the quarks and gluons, while the coupling $g(T)$ is obtained fitting lQCD thermodynamics \cite{Borsanyi:2010cj,Borsanyi:2016ksw}.
On one hand, the advantage of this approach consists in the understanding the degrees of freedom of QCD at finite temperature. On the other hand, a microscopic description of the plasma has the advantage to be suitable for microscopic simulations, based on transport theory, whose aim is the description of the fireball created in ultra-relativistic Heavy Ion Collisions \cite{Scardina:2012mik,Ruggieri:2015tsa,Scardina:2017ipo,Plumari:2019gwq,Sambataro:2022sns}. Other approaches, such as T-Matrix model, hint at quasi particle description with temperature dependent masses for $T> 1.5$ $T_c$, but closer to $T_c$ and for low momenta $p\sim T$, they find spectral functions that are quite broader, challenging a description in terms of quasi-particle for gluon and light quarks. \cite{Liu:2016ysz, Liu:2017qah}.
Instead, as far as the heavy flavour sector is concerned, in the past years the $QPM$ has led to a good description of the main observables of $D$ mesons \cite{Oliva:2020doe, vanHees:2005wb,vanHees:2007me,Gossiaux:2008jv,Alberico:2011zy,Uphoff:2012gb,Lang:2012nqy,Song:2015sfa,Das:2013kea,Cao:2015hia,Das:2015ana,Cao:2017hhk,Das:2017dsh,Sun:2019fud,Coci:2019nyr,Li:2019wri,Cao:2018ews,Rapp:2018qla,Ravagli:2007xx}. The QPM has also been extended to the study of bottom dynamics, predicting $R_{AA}(p_T)$ and $v_2(p_T)$ in agreement within current error bars to the ALICE data on semi-leptonic decay \cite{Sambataro:2023tlv}. This leads to an extrapolation of the spatial diffusion coefficient $D_s(T)$ of Heavy-Quarks (HQs) in agreement with the available lQCD calculations \cite{Scardina:2017ipo,Plumari:2019hzp,Sambataro:2022sns,Sambataro:2023tlv}.
However, one has to consider that such an agreement is quite quantitatively approximate due to the inherent uncertainties in both lQCD data and the phenomenological extrapolation based on experimental data, with significant error bars especially in the low momentum region. Also a potential role of an initial Glasma phase \cite{Sun:2019fud} or of non-Markovian dynamical effects \cite{Pooja:2023gqt} allow a comprehensive understanding of both pA and AA observables. Although the $QPM$ was successfully applied to describe lQCD results on the EoS with $N_f=2+1$ \cite{Plumari:2011mk}, the large "thermal average" of quark masses obtained in the simple $QPM$ leads to an underestimation of the quark number susceptibilities $\chi_q$ of light quarks, casting some doubts on the capability of this approach to correctly describe the inner dynamical structure of lQCD \cite{Plumari:2011mk}. In order to simultaneously describe the EoS of lQCD and the quark number susceptibilities, we study the possibility to consider 
massive quasi-particles with partonic propagators which explicitly depend on the three-momentum wrt the medium, such that they reach the current quark masses of perturbative QCD (pQCD) at high momenta, following the idea developed in Ref. \cite{Berrehrah:2015vhe, Berrehrah:2016vzw}. Furthermore, we also extend our approach to the new lattice results for the Equation of State of QCD with $2+1+1$ dynamical flavours (i.e. including also the charm quark in the bulk) and we estimate the charm spatial diffusion coefficient $D_s(T)$.
\newline 
The paper is organized as follows: in Section II, we remind the main properties of Quasi-Particle Model with $N_f=2+1$ as developed in \cite{Plumari:2011mk} and we introduce the analogous approach with $N_f=2+1+1$, which includes the charm quark. In Section IIa, we discuss the main extension of the $QPM$ proposed in this paper, in which quark and gluon masses become functions of both temperature $T$ and momentum $p$, following the model developed by the PHSD group \cite{Berrehrah:2015vhe, Berrehrah:2016vzw}. The quark number susceptibilities for light, strange and charm quarks, are discussed in Section IIb, also in relation to different T parametrizations for the charm quark mass $m_c(T)$. In Section III, we discuss the spatial diffusion coefficient $D_s(T)$ for the charm quark, making a comparison between the simple $QPM$ and the extension $QPM_p$ and also evaluating the infinite mass limit to correctly make a comparison with lQCD calculations. In the last Section, we show our study on the evolution of the charm quark phase-space distribution function in the different approaches, in terms of nuclear modification factor $R_{AA}$.

\section{The Quasi Particle Model}

\begin{figure}
\centering
\includegraphics[scale=0.3]{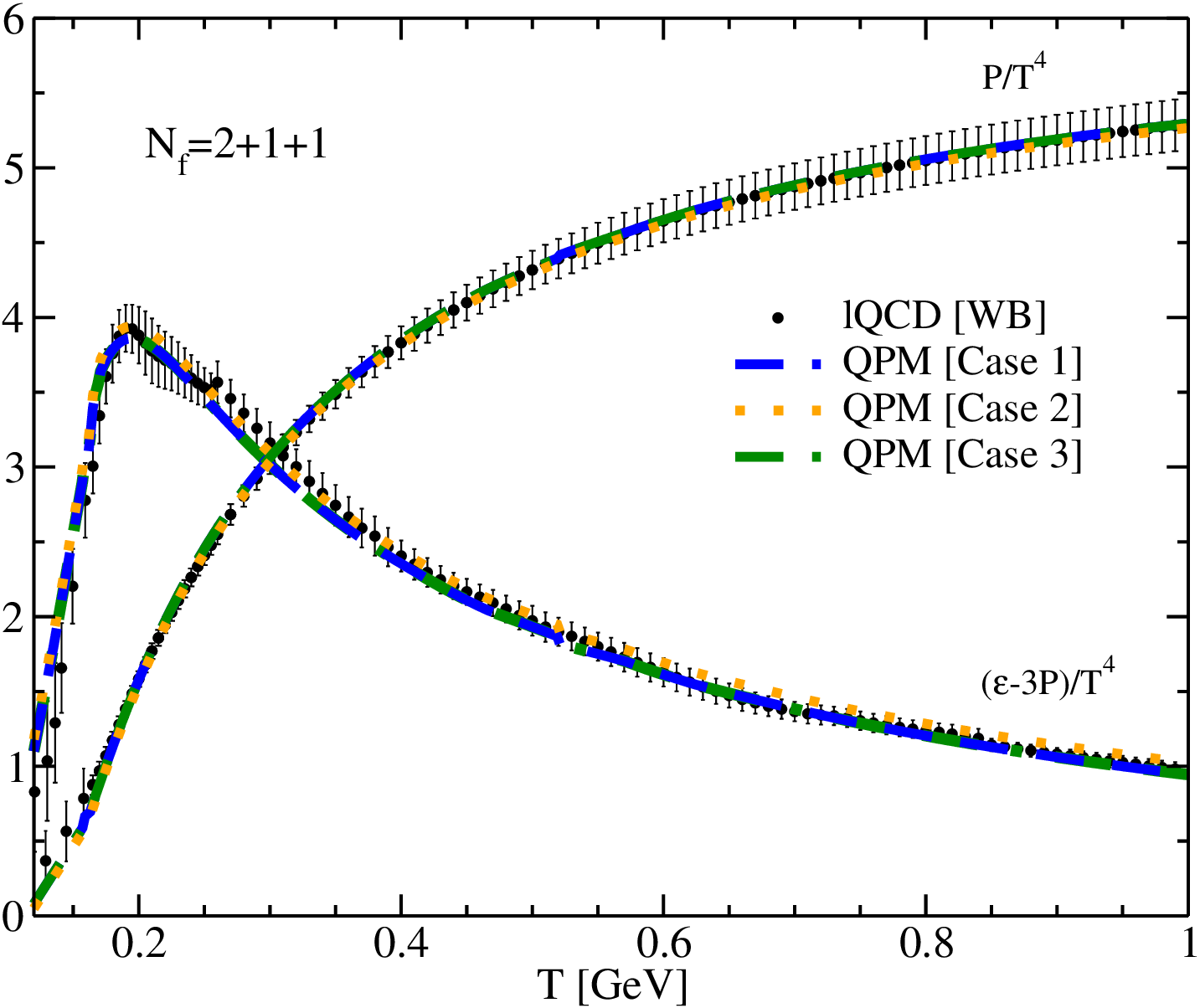}
\caption{Comparison between scaled pressure $P/T^4$ and interaction measure $I=(\epsilon - 3P)/T^4$ obtained in $QPM$ approach and lQCD with $N_f=2+1+1$. Lattice QCD data taken from Wuppertal-Budapest collaboration \cite{Borsanyi:2016ksw}. Different curves are for the different cases considered: blue dashed line for Case 1, orange dotted line for Case 2 and green dot-dashed line for Case 3.}\label{fig:2}
\end{figure}

It is well known that the $QPM$ approach gives the possibility to quantitatively describe lQCD Equation of State even in the vicinity of phase transition, where pQCD calculations fail \cite{Plumari:2011mk}. In this way, the model 
provides the possibility to account for non-perturbative dynamics by encoding the interaction in the quasi-particle masses. 
In $QPM$ the quark and gluon masses are related to the temperature and coupling $g(T)$ by the following standard pQCD relations \cite{Levai:1997yx, Peshier:2005pp, Peshier:1995ty}: 
\begin{eqnarray}\label{mass_QPM}
&&m_g^2=\frac{1}{6}g(T)^2\left[\left(N_c+\frac{1}{2}N_f\right)T^2+\frac{N_c}{2\pi^2}\sum_{q}\mu_q^2\right],\nonumber \\
&&m_{u,d}^2=\frac{N_c^2-1}{8N_c}g(T)^2\left[ T^2+\frac{\mu^2_{u,d}}{\pi^2}\right]
\end{eqnarray}
where $N_c$ is the number of colors, $N_f$ is the number of flavors considered, $m_{u,d}$ are the masses of the light quarks and $\mu_q$ is the chemical potential of the generic flavor $q$ considered. In this paper, we will consider $N_f=3$ and $N_f=4$ for the extension including the charm quark, as will be discussed later.
Following Ref.\cite{Plumari:2011mk}, we have considered the following temperature dependence for the strange quark mass:
\begin{equation}\label{mass_s}
m_{s}^2-m_{s0}^2=\frac{N_c^2-1}{8N_c} g(T)^2
\left[ 
T^2+\frac{\mu^2_{s}}{\pi^2} 
\right]
\end{equation}
where $m_{s0}=0.150 \, GeV$ is the finite current quark mass. 
\newline
In our approach the coupling $g$ is a temperature-dependent function which has to be determined through the fit to lQCD data.
The quasi-particle quarks and gluons behave like a free massive gas and the total pressure of the system can then be written as the sum of independent contributions coming from the different constituents. However, when the  temperature-dependent masses of Eq.s \eqref{mass_QPM} and \eqref{mass_s} are included in the pressure, its derivative with respect to the temperature will produce an extra term in the energy density which does not have the ideal gas form. The model is therefore completed by introducing a temperature-dependent term that accounts for further non-perturbative effects: the pressure and consequently the energy density will contain an additional contribution, called bag constant $B(T)$, introduced with the purpose of ensuring thermodynamic consistency. That being said, the total pressure of the system can be written in the following form:
\begin{equation}\label{pressure}
P_{qp}(m_i,...,T)=\sum_{i=u,d,s,c,g}d_i\int\frac{d^3 p}{(2 \pi)^3}\frac{\textbf{p}^2}{3 E_i(p)}f_i(p)-B(T)
\end{equation}
where $f_i(p) = [1 \mp \exp (\beta E_i(p))]^{-1}$ 
are the Bose and Fermi distribution functions, with $E_i(p) =\sqrt{\textbf{p}^2+m_i^2}$, $d_i$ = 2 $\times$ 2 $\times$ $N_c$ for quarks and $d_i = 2 \times (N^2_c-1)$ for gluons. The thermodynamic consistency is satisfied by the following set of equations: 	
\begin{equation}\label{BAG}
\frac{\partial B}{\partial m_i}+d_i\int\frac{d^3p}{(2\pi)^3}\frac{m_i}{E_i}f_i(E_i)=0.
\end{equation}
Finally, imposing thermodynamic consistency and using the thermodynamic relationship between energy density and pressure $\epsilon(T) = TdP(T)/dT - P(T)$, we obtain that the energy density can be written as:
 \begin{eqnarray}
\epsilon_{qp}(T)&=& \sum_i d_i \int \frac{d^3p}{(2\pi)^3}E_if_i(E_i)+B(m_i(T)) \nonumber\\&=&
\sum_i\epsilon_{kin}^i(m_i,T)+B(m_i(T)). \label{energy}
\end{eqnarray}
In this approach the coupling constant $g(T)$ in Eq.\eqref{mass_QPM} is not known, to find it we match the energy density obtained within lQCD data with the energy density from $QPM$ in Eq.\eqref{energy}:
\begin{equation}\label{en_lQCD}
\epsilon_{qp}(T)= \epsilon_{lQCD}(T).
\end{equation}
Notice that the $QPM$ effective coupling extracted will be larger than the one obtained in pQCD calculations, especially when the temperature of the system approaches the critical value $T_c=0.155$ $GeV$. For more details about $QPM$ in $N_f=2+1$, see Ref.\cite{Plumari:2011mk}.
In the following, we show the $QPM$ extension which includes charm quarks and the main differences obtained wrt the $N_f=2+1$ case. The extension to the $N_f=2+1+1$ case is obtained by fitting the available lQCD energy density data from the Wuppertal-Budapest (WB) collaboration in Ref.\cite{Borsanyi:2016ksw}, which include the charm quark in the bulk.
In particular, we want to discuss the role of different temperature dependencies for the charm mass in the $QPM$ and, furthermore, the impact of momentum dependence in parton masses on EoS, quark susceptibilities and heavy quarks transport coefficients.
In this section, we concentrate on temperature dependence and in the next section we discuss the effect of the momentum dependence. For what concerns the heavy quark mass, we have considered the following three cases, all corresponding to a quite good description of the energy density and pressure (see Fig.\ref{fig:2}):\\

\textit{Case 1}:
A constant mass for the charm quark $m_c=1.5$ $GeV$.
\\

\textit{Case 2}: A temperature dependent charm quark mass, with thermal mass assumed to be given by the same parametrized form of the strange quark in the $QPM$, i.e. $m_c^2=m_{c0}^2+\frac{N_c^2-1}{8 N_c}g^2 [T^2+\frac{\mu^2_c}{\pi^2}]$ with $m_{c0}=1.3$ $GeV$.
\\

\begin{figure}[h]\centering
\includegraphics[scale=0.28]{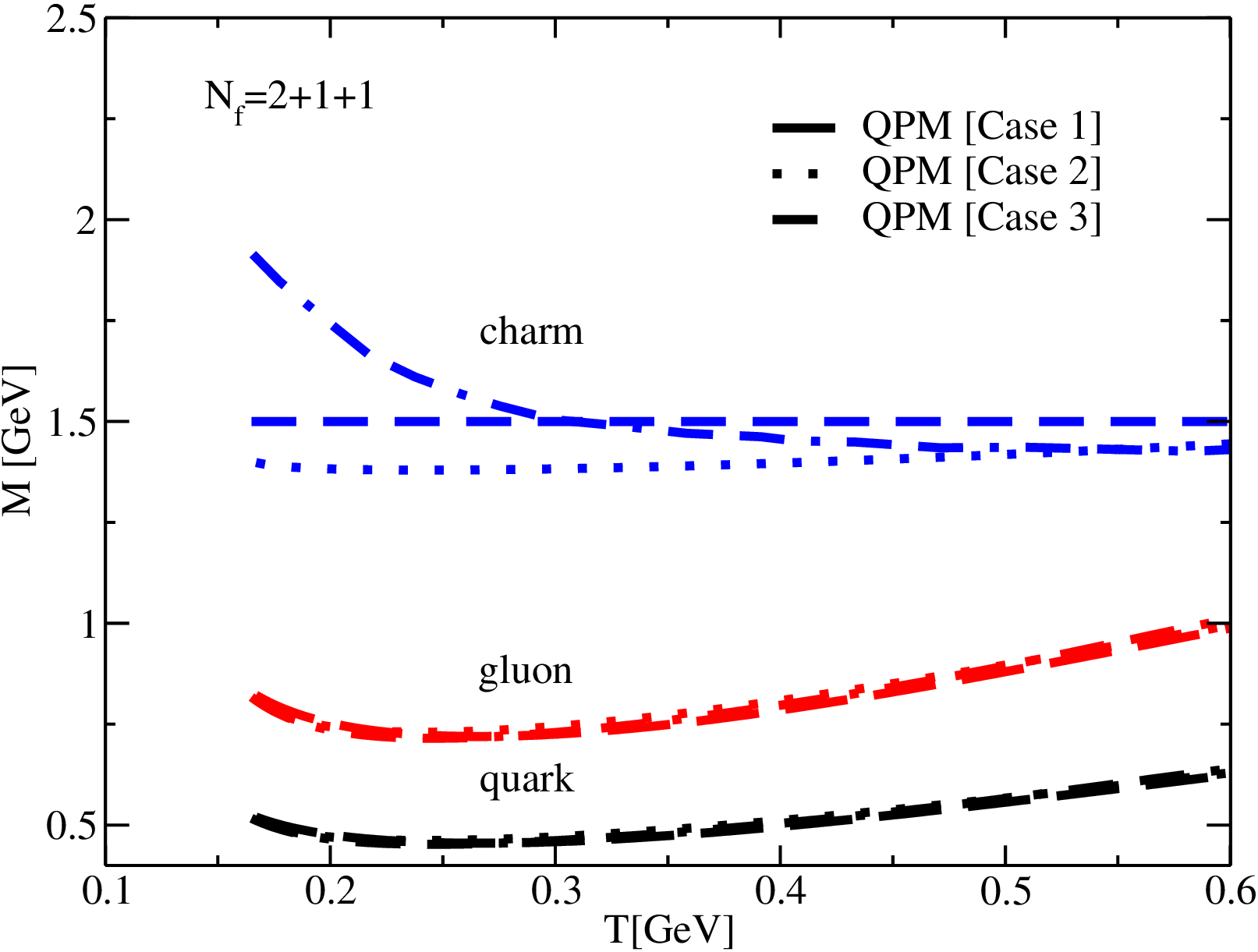}
\caption{Gluon (red lines), light quark (black lines) and charm quark (blue lines) masses as a function of the temperature.
Different styles are for the three different cases considered: dashed line for Case 1, dotted line for Case 2 and dot-dashed line for Case 3.
}\label{fig:3}
\end{figure}

\textit{Case 3}:
The in-medium mass of charm quark is constrained by charm number fluctuations. In particular, in this case the $T$-dependent charm quark mass has been extracted from the charm susceptibility, $\chi_2^c=\frac{T}{V}\frac{\partial^2 \ln Z}{\partial \mu_{i}^2}$, following the work presented in Ref.s \cite{Bazavov:2014yba, Rapp:2018qla, Petreczky:2009cr}. The following expression for the quark fluctuations:
\begin{equation}\label{case 3}
c_2^q=\frac{\chi_2^q}{2}=\frac{1}{2}\frac{6}{\pi^2}\left(\frac{m_q}{T}\right)^2 \sum_{l=1}^\infty (-1)^{l+1} K_2 (lm_q/T)
\end{equation}
can be solved in terms of $m_q/T$, with the $\chi_2^q$ values numerically obtained in lQCD from Ref. \cite{Bellwied:2015lba}.
We then fit the resulting temperature dependence of charm mass, which we show in Fig. \ref{fig:3} as blue dot-dashed line. Notice that this case reproduces by construction the charm quark susceptibility (see Fig. \ref{fig:10}) which we will discuss the next section.\\
\begin{figure}[ht!]
\centering
\includegraphics[scale=0.3]{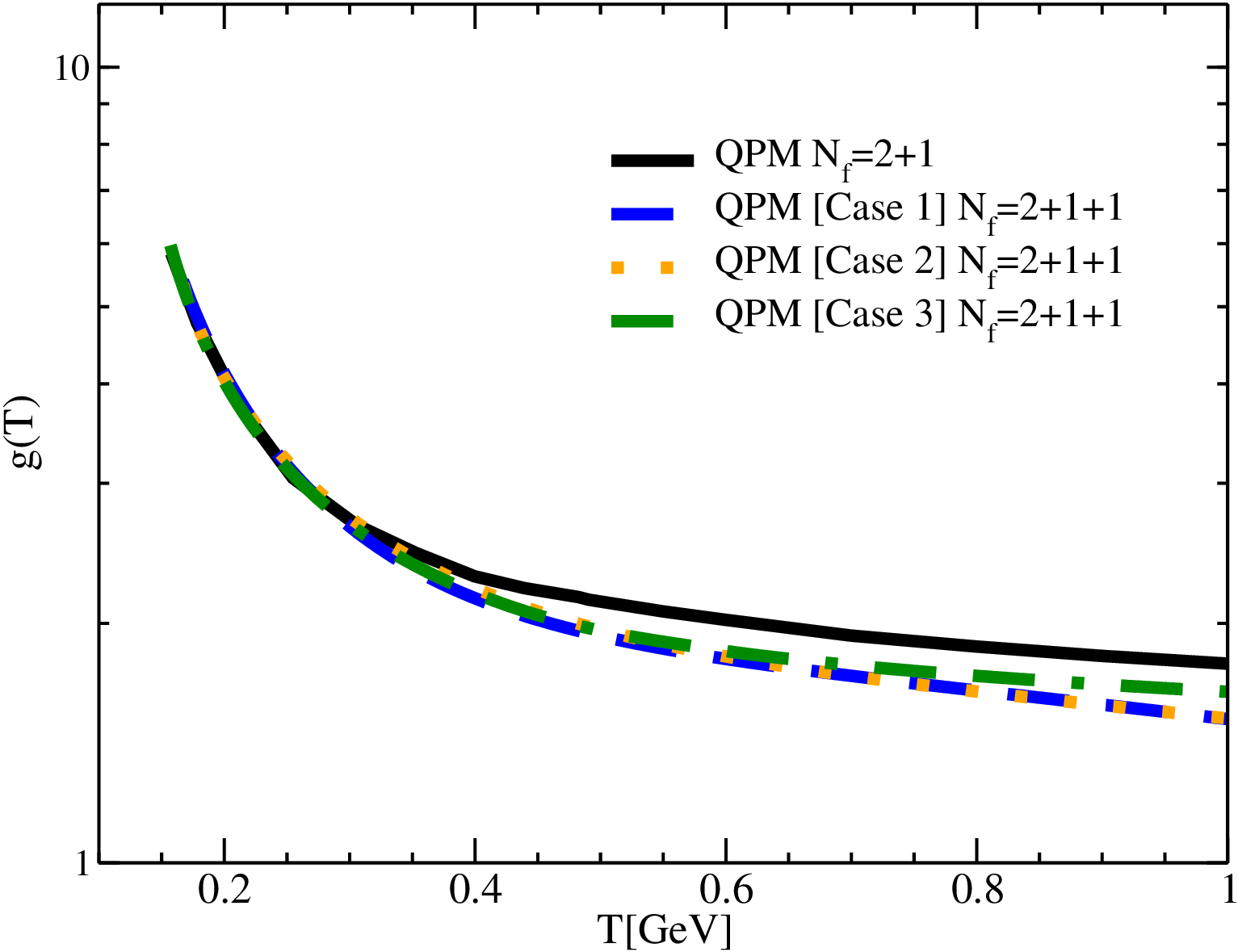}
\caption{Effective coupling $g(T)$ as a function of the temperature. Different lines correspond to the different cases described in the text for the extension to $N_f=2+1+1$: blue dashed line for Case 1, orange dotted line for Case 2 and green dot-dashed line for Case 3. The black solid curve corresponds to the standard $QPM$ result for $N_f=2+1$.
 }\label{fig:1}
\end{figure}
The resulting $g(T)$ that we extract from the procedure discussed above is shown in Fig.\ref{fig:1}. The black solid curve describes the coupling in the $QPM$ with $N_f=2+1$ from Ref. \cite{Plumari:2011mk} and the other curves describe the $g(T)$ obtained within the $QPM$ with $N_f=2+1+1$ for the different parametrizations of the charm mass. The temperature behaviour of the effective coupling $g(T)$ in the region $T < 2$ $T_c \approx 0.3 \, GeV$ is almost the same for all cases and similar also to the one with $N_f=2+1$ flavours, which does not include the charm quark in the medium. However, we notice that all the curves for $N_f=2+1+1$ stand lower with respect to the $N_f=2+1$ case for $T>2T_c$ of about $15$ $\%$, a discrepancy which mainly comes from the difference in the energy density of the two systems as revealed in the lQCD data. The three cases with $N_f=2+1+1$ do not show significant differences in the temperature dependence of the corresponding $g(T)$ in all temperature regions shown in figure.
Once we have obtained both the coupling $g(T)$ and the bag $B(T)$ (by fitting the lattice data for the energy density), we can calculate the other thermodynamic quantities like pressure and trace anomaly $I=\epsilon - 3p$. In Fig. \ref{fig:2}, we show the good agreement between lQCD data with $N_f=2+1+1$ \cite{Borsanyi:2016ksw} for pressure and trace anomaly and our results that we obtain for all the different cases as explained above.
The corresponding masses of plasma particles as function of temperature obtained within $QPM$ approach by Eq.\eqref{mass_QPM} are shown in Fig.\ref{fig:3}. In the plot, the black lines are for the light quarks and the red lines for the gluons, while the blue curves describe the temperature dependence of charm quark in the three temperature parametrizations considered.
Since the resulting $g(T)$ is similar for all cases, the corresponding light quark and gluon masses have very similar behavior as function of T. Regarding the charm quark mass, there is a relevant difference in the temperature dependence, especially for \textit{Case 3} which at $T\approx T_c$ assumes quite large values ($m_c\approx 1.9$ $GeV$). The trend we find is very similar to the one already reported in Ref. \cite{Petreczky:2009cr}, where the susceptibility $\chi_c(T)$ is fitted in a quasi-particle picture.

\subsection{Extension of QPM to momentum dependent quasi-particle masses}
In this section we study the possibility to extend the conventional $QPM$, with quasi-particle masses that depend only on temperature $m_{g,q}(T)$, to a more realistic model in which the quasi-particle masses acquire a momentum dependence. This extension has the aim to allow the masses to decrease toward the current ones, mimicking a perturbative behaviour at high momenta and it is motivated by the fact that the simple $QPM$ does not take into account the possibility that when the quasi-particle momentum is large compared to the temperature of the system, this should lead to partons with current quark masses. In fact, in the standard $QPM$ approach, the coupling $g(T)$ is fixed by a fit to lQCD energy density, implying large thermal masses that account for the non perturbative dynamics. However, such an approach misses completely the evolution of the masses wrt momentum: that is naturally expected as a consequence of asymptotic free dynamics and it is known to be present also at zero temperature for the chiral condensate or in many-body approaches like T-matrix or Dyson-Schwinger \cite{Liu:2017qah, ZhanduoTang:2023ewm, Fischer:2003rp, Fischer:2006ub, Fischer:2014ata}. The approach used in this work follows the same model developed by the PHSD group in the ref. \cite{Berrehrah:2015vhe} where the momentum dependence of the quark mass is motivated by Dyson-Schwinger studies \cite{Fischer:2006ub, Mueller:2010ah}: in particular, the propagator in the low momentum region behaves like the propagator of a massive particle and reduces to the bare perturbative propagator as momentum increases. We can express this in a simplified way with a momentum-dependent thermal mass which drops from the effective value defined by the fit to lQCD energy density to the chiral mass (or gluon condensate) for light quarks (or gluons) with $|p| \rightarrow \infty$. 
In the text we will refer to $QPM_{p}$ as the extended version of $QPM$ where we include such momentum dependent quasi-particle masses. We will show that within this extended $QPM_p$, it is still possible to reproduce the lQCD equation of state at finite temperature T, but it will imply larger quark number susceptibilities wrt $QPM$ in agreement with the available lQCD data by WB collaborations \cite{Borsanyi:2016ksw}. 
The quasi-particle masses are expressed as a function of temperature $T$ and squared momentum $p^2$ by the following relations taken from Ref. \cite{Berrehrah:2015vhe}:
\begin{eqnarray}\label{gluon_p_mass}
M_g(T,p)&=&\frac{3}{2}\sqrt{\frac{g^2(T)}{6}\left[\left(N_c+\frac{N_f}{2}\right)T^2\right]}\cdot h(\lambda_g,p)+m_{\chi g} \nonumber	\\
M_{q}(T, p)&=&\sqrt{\frac{N_c^2-1}{8N_c}g^2(T) T^2}\cdot h(\lambda_q,p)+m_{\chi q} 
\end{eqnarray}
where the new element is the momentum-dependent factor $h(\lambda,p)$ in the masses, which guarantees the pQCD limit to be reached for $p \rightarrow \infty$: 
\begin{equation}\label{mom_factor}
h(\lambda_{q,g},p)=\frac{1}{\sqrt{1+\lambda_{q,g} (T_c/T)^2 \, p^2}}
\end{equation}
with the coefficient $\lambda_{q,g}$ which is given by  $\lambda_g=5 \, GeV^{-2}$ for gluons and $\lambda_q=12 \, GeV^{-2}$ for quarks. In the above expressions, $m_{\chi g} \approx 0.5 \, GeV$ is the gluon condensate and $m_{\chi q}$ is the light quark chiral mass which is $m_{\chi q}=0.003 \, GeV$ for up and down quarks and $m_{\chi q}=0.06 \, GeV$ for the s quark: these ensure the right limit for $|p| \rightarrow \infty$.\\

We have studied the possibility to employ a momentum dependence in charm quark mass following the functional form of Ref. \cite{Fischer:2006ub}, but this momentum dependence, as can be expected, does not provide significant change in EoS and effective coupling $g(T)$. In our calculations, we therefore neglect this additional factor and consider a momentum independent mass for charm quark.\\
In the calculation referred as $QPM_{p}$ [Case 1], the charm quark mass is fixed to $m_c=1.5$ $GeV$. 
However, we will discuss in the next section the role of a temperature dependence for the charm quark mass in both $QPM$ and $QPM_p$ and its effect on the spatial diffusion coefficient $D_s$. 
\begin{figure}[ht!]
\centering
\includegraphics[width=0.95\linewidth]{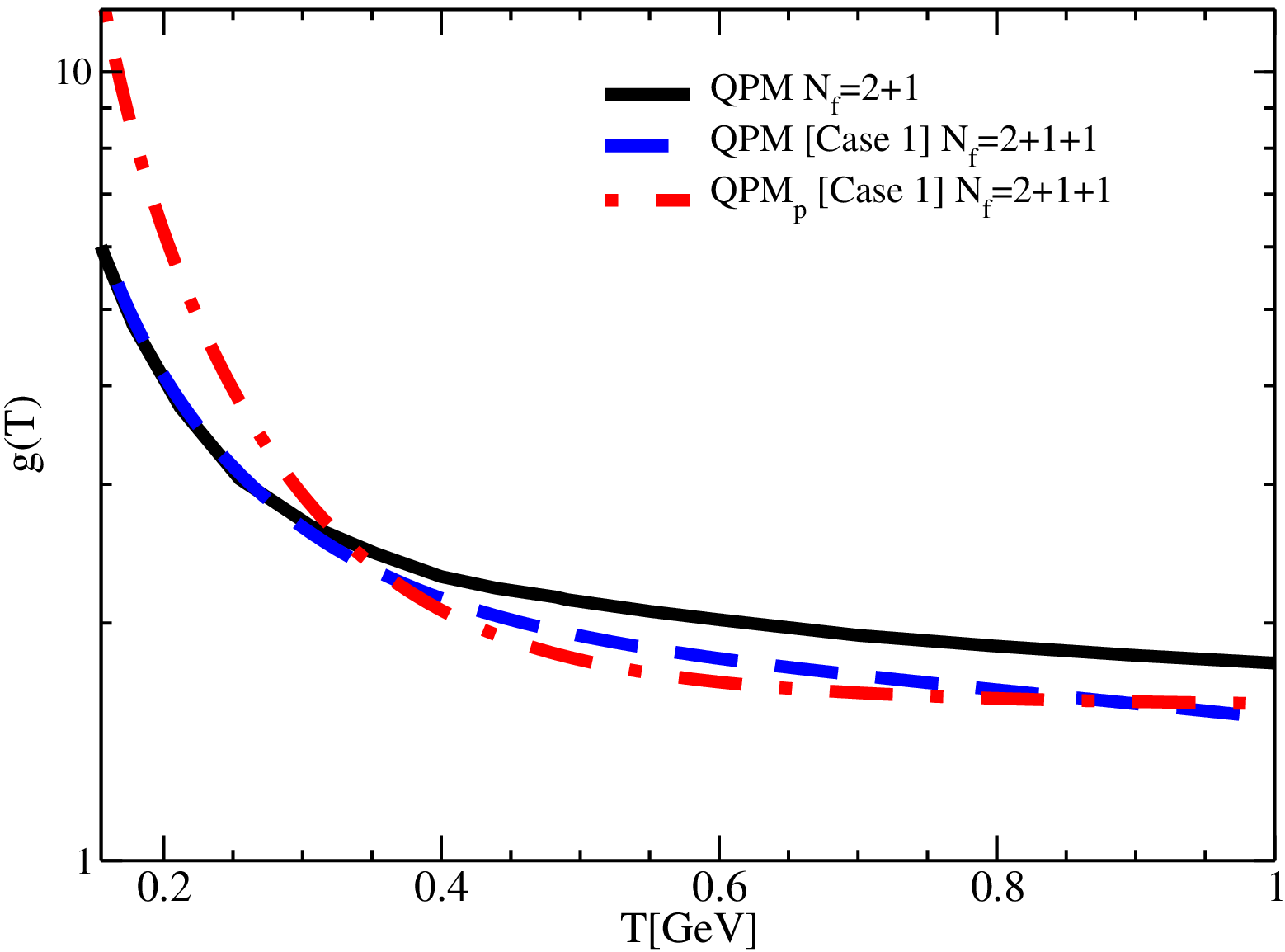}
\caption{Comparison between $g(T)$ for $N_f=2+1+1$ obtained in standard $QPM$ with $m_c=1.5$ $GeV$ [Case 1] and $QPM_{p}$ with $m_c=1.5$ $GeV$ [Case 1]. The black solid line is the result for $QPM$ with $N_f=2+1$ flavours.}\label{fig:4}
\end{figure}
In a similar way done for the standard $QPM$, we extract the coupling $g(T)$ for the $QPM_{p}$ using the matching condition in Eq.\eqref{en_lQCD} to reproduce the energy density. The temperature dependence of the coupling $g(T)$ in $QPM_{p}$ is shown in Fig.\ref{fig:4} in comparison with the 
coupling extracted for the standard $QPM$ for both $N_f=2+1$ and $N_f=2+1+1$.
Notice that the $g(T)$ in $QPM_{p}$ is considered only as a function of the temperature as in the standard $QPM$ even if, a more complete and microscopic approach, should take into account the evolution with momentum of the coupling: this more realistic extension is postponed to future works.
As shown in Fig.\ref{fig:4}, near $T_c$ the effective coupling $g(T)$ in the $QPM_{p}$ is larger than the one in standard $QPM$ of about a factor $2$. Instead, at high temperatures we find that the $QPM_{p}$ result becomes similar to the result from standard $QPM$ with $N_f=2+1+1$. The behaviour of $g(T)$ in the low T region is motivated by the fact that the parton masses in $QPM_{p}$ decrease as function of momentum so the effective coupling $g(T)$ that determines the mass $m(T,p\rightarrow 0)$, has to become larger to reproduce the same energy density that is dominated by the smaller mass at finite $p$. On the other hand, in the high temperature region the energy density becomes less sensitive to the momentum dependence of the quasi-particle masses and the coupling $g(T)$ from $QPM_{p}$ plays a similar role to the standard $QPM$.
In general, the effective coupling $g(T)$ obtained numerically for the cases considered in the paper, can be expressed by the following parametrization:
\begin{equation}
  g^2(T)=\frac{48 \pi^2}{(11N_c- 
2N_f)\ln\left[\alpha (\frac{T}{T_c})^2+\lambda\frac{T}{T_c}+\frac{T_s}{T_c}\right]^2}
\label{Eq:coupling_new}
\end{equation}
where $N_c$ and $N_f$ are respectively the number of color and flavors and the parameters $\alpha$, $\lambda$ and $T_s/T_c$, obtained by fitting the $g(T)$ for the different cases considered, are shown in Table \ref{table:param}. Notice that the fits of Eq. \eqref{Eq:coupling_new} are valid up to $T\sim 4T_c$.
\begin{table} [ht]
\begin{center}
  \begin{tabular}{l |c c c }
    \hline
    \hline
      &$\alpha$ & $\lambda$ &$T_s/T_c$ \\ 
      \hline
    $QPM$ $N_f=2+1$     & 0  & 2.6 & -0.57 \\
    $QPM$ $N_f=2+1+1$   & 0   & 3.3 & -1.37  \\
    $QPM_{p}$ $N_f=2+1+1$   & 1.1   & -0.12 & 0.37 \\
\hline
\hline
\end{tabular}
\end{center}
\caption{Fitted parameters $\alpha$, $\lambda$ and $T_s/T_c$ for the different cases considered in the text. 
}\label{table:param}
\end{table}

\begin{figure}[ht!]\centering
\includegraphics[width=0.9\linewidth]{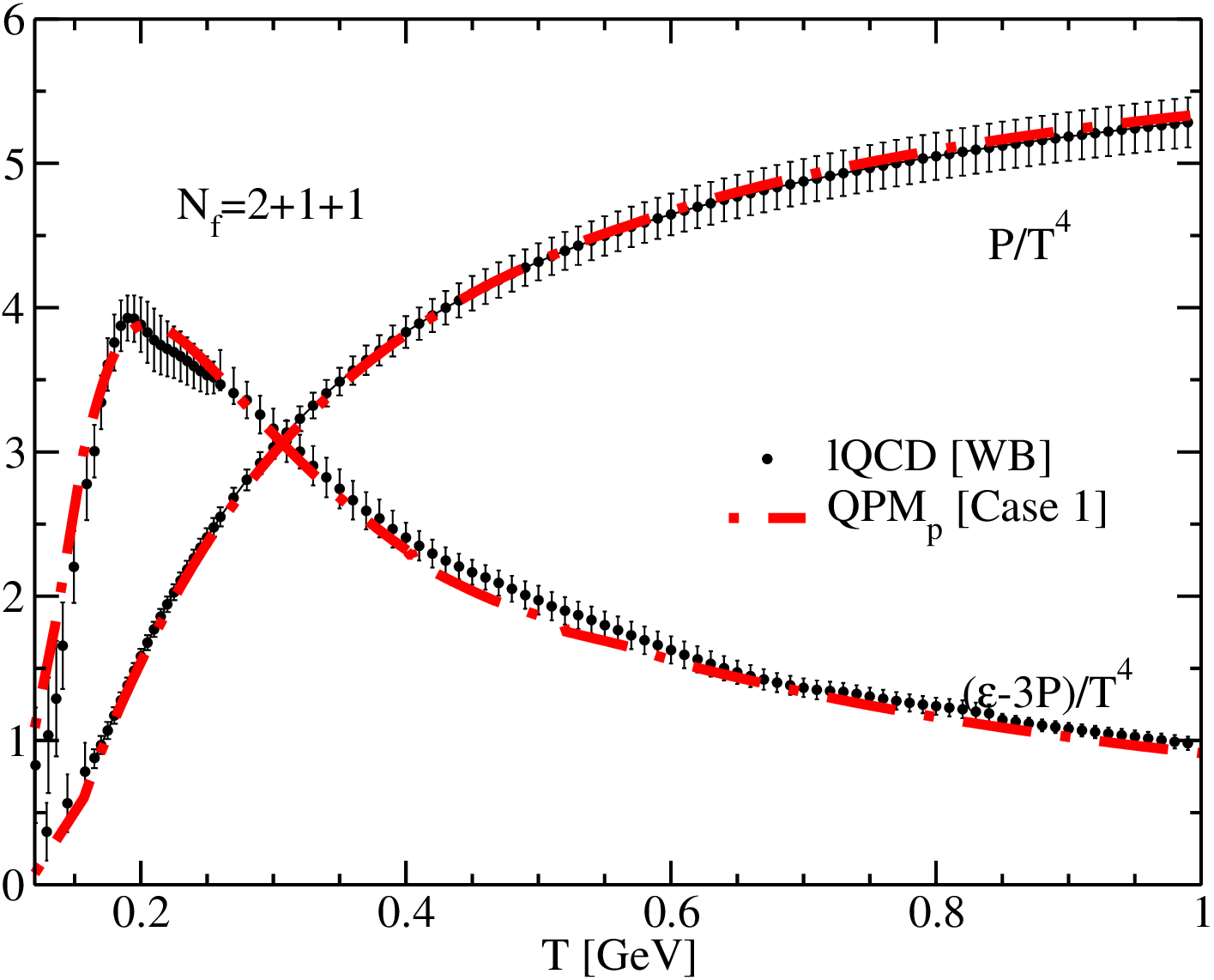}
\caption{Scaled pressure $P/T^4$ and interaction measure $I=(\epsilon - 3P)/T^4$ obtained in $QPM_{p}$ (red dot-dashed lines) and lQCD data taken from Ref.\cite{Borsanyi:2016ksw}. }\label{fig:5}
\end{figure}

In Fig.\ref{fig:5}, we compare the results for $QPM_{p}$ of scaled pressure and interaction measure with lattice QCD calculations taken from Ref.\cite{Borsanyi:2016ksw}. We notice that our results are in a quite good agreement with the lattice data.
\begin{figure}[ht!]\centering
 \includegraphics[width=1.0\linewidth]{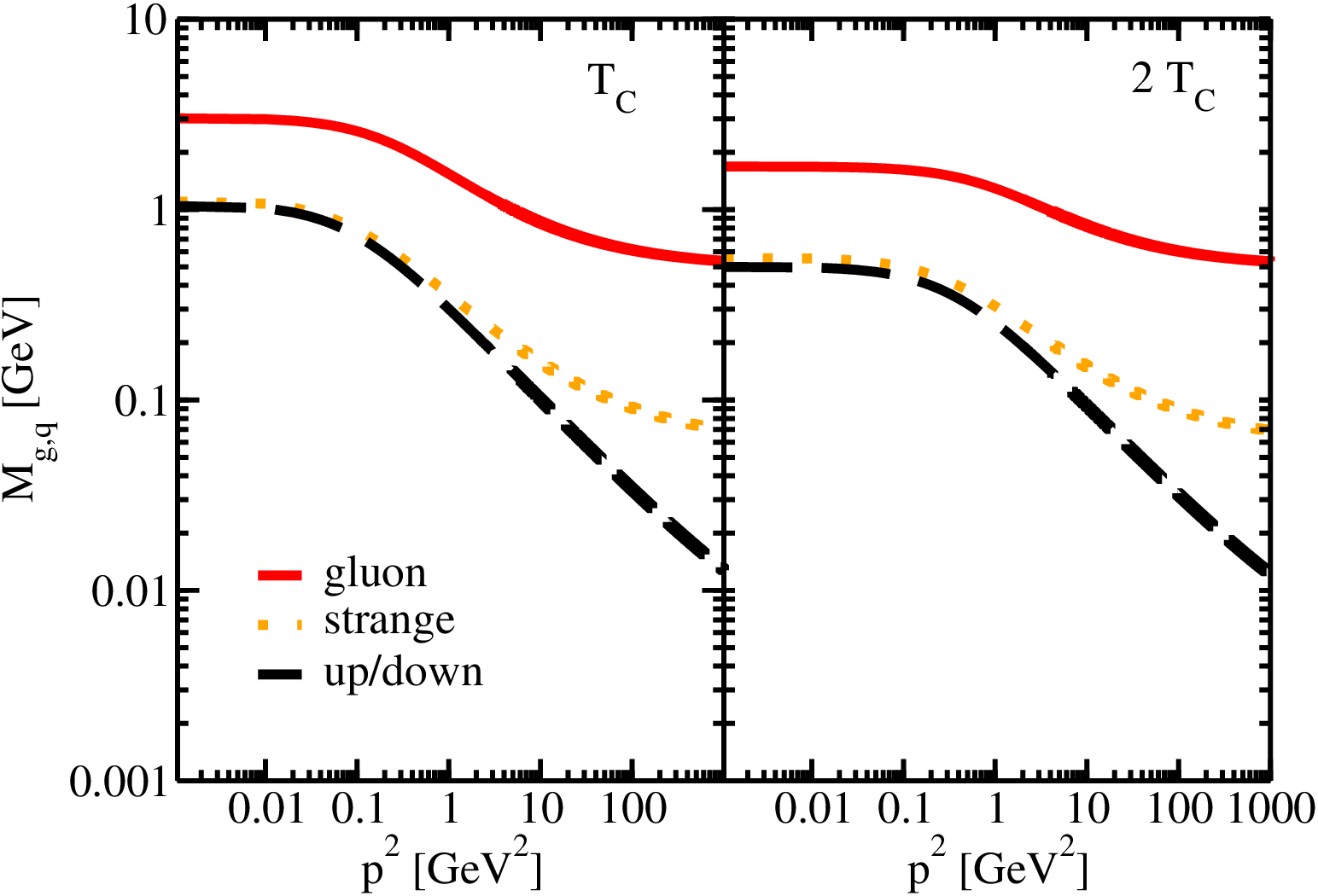}
 \caption{Quasi-particle masses as function of squared momentum $p^2$ in the $QPM_{p}$ for two different temperatures $T=T_c=0.155 \, GeV$ (left panel) and $T=2T_c=0.310 \, GeV$ (right panel). Red solid lines are for gluons, orange dashed for strange quarks and black dashed lines for light quarks.}\label{fig:6}
\end{figure}	
Once one obtains the coupling $g(T)$, the corresponding quasi-particle masses in $QPM_{p}$ can be calculated using Eq.\eqref{gluon_p_mass}.
In Fig.\ref{fig:6}, we show the quasi-particle masses as function of $p^2$ for two different values of the temperature $T=T_c$ (left panel) and $T=2 \, T_c$ (right panel). As shown, we recover the correct pQCD limit at very high momenta, where the masses approach the current quark masses.
\begin{figure}[ht]\centering
\includegraphics[width=1.0\linewidth]{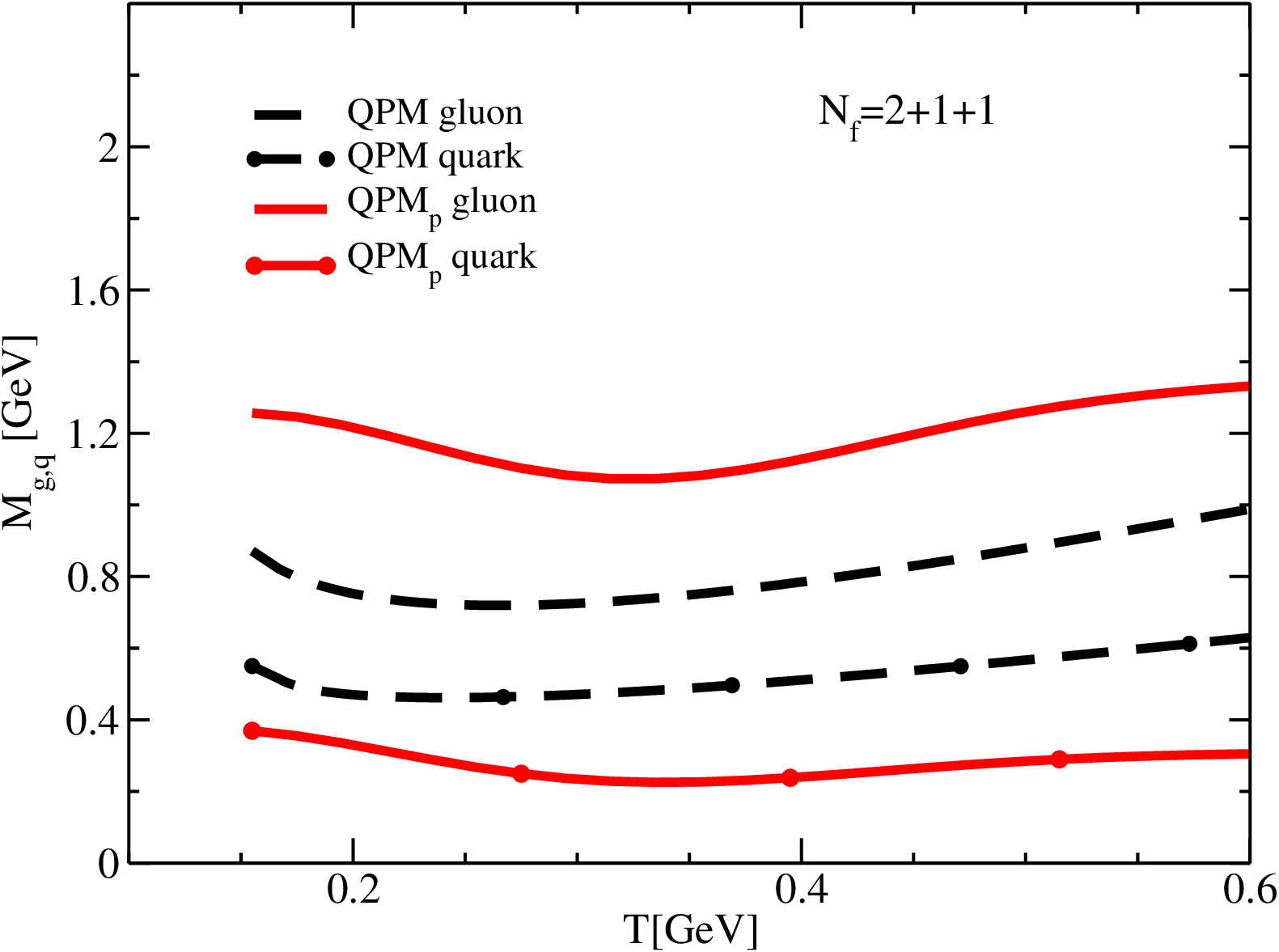}
\caption{Quasi-particle masses of gluon and light quarks as a function of the temperature in both $QPM$ and $QPM_p$. The masses in $QPM_{p}$ have been obtained considering the mean value of momentum at each value of temperature (see main text)}.\label{fig:7}
\end{figure}
In Fig.\ref{fig:7}, the corresponding temperature dependence of the quasi-particle masses of up and down quarks and gluons in $QPM_p$ is shown in comparison with the masses obtained in $QPM$ approach without momentum dependence. In particular, the $QPM_p$ masses are plotted at the mean thermal value of particle momentum:

\begin{equation}
  \langle p\rangle(T)=\frac{\int_0^\infty |\Vec{p}| f_i(\Vec{p},T) d^3 p}{\int_0^\infty f_i(\Vec{p},T) d^3 p}
\label{Eq:mean_mom}
\end{equation}

for each value of temperature. We notice that the ratio $m_g/m_q$ in the extended approach is larger by a factor $2$ wrt the standard $QPM$ in the whole temperature region. This is due to the fact that at large momentum $p$ the quark mass goes to zero while for the gluon mass even in $p\rightarrow \infty$ limit, we have the gluon condensate value $m_{\chi g}=0.5$ $GeV$. One may also notice that light quark have masses between $300$ $MeV$ and $380$ $MeV$ in all the temperature range considered. This also implies that $QPM_p$ has a microscopic structure where light quarks have larger density than gluons with respect to $QPM$ where the quark-gluon difference is smaller. Despite this, the diffusion coefficient that we will discuss in the next section has similar or even smaller value wrt $QPM$.

\subsection{Susceptibilities}

The quark number susceptibilities are encoded in the lQCD equation of state at finite but small chemical potential over temperature ratios ($\mu_q/T\ll 1$) but provide useful information about the nature of the degrees of freedom close the critical temperature $T_c$ \cite{Bazavov:2009zn, Bellwied:2015lba}. The standard quasi-particle models, which are tuned to reproduce the lQCD equation of state, tend to underestimate the susceptibility \cite{Plumari:2011mk}.
We conclude this section by showing the predictions on quark number susceptibilities obtained within the three different cases for the charm mass behaviour of $QPM$ used and within its extension to the case with momentum dependent quasi-particle masses ($QPM_{p}$).
In particular, we show predictions for light, strange and charm quark number susceptibilities which are expressed by:
\begin{equation}\label{formula_susce}
\chi_{u,s,c}=\frac{T}{V}\frac{\partial^2 \ln Z}{\partial \mu_{u,s,c}^2} 
\end{equation}
\begin{figure}[t!]
\centering
\includegraphics[width=1.0\linewidth]{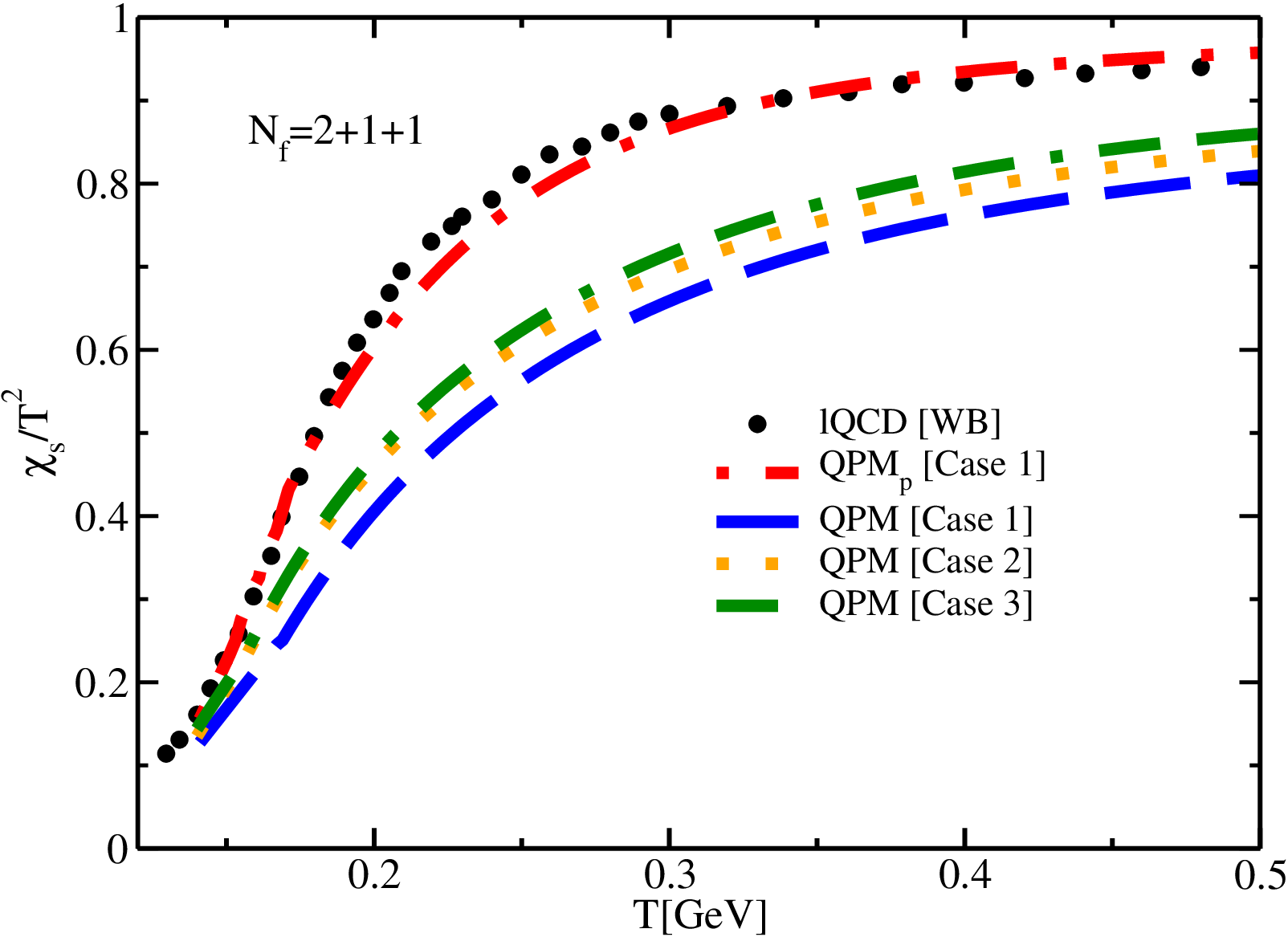}
\caption{Strange quark number susceptibilities $\chi_s$ as functions of the temperature, for the different charm quark mass parametrizations, in the standard $QPM$ and for the extended $QPM$ model with momentum dependent quasi-particle masses.
The lQCD data have been taken from Ref.\cite{Bellwied:2015lba}.
}\label{fig:8}
\end{figure}
\begin{figure}[t!]
\centering
\includegraphics[width=1.0\linewidth]{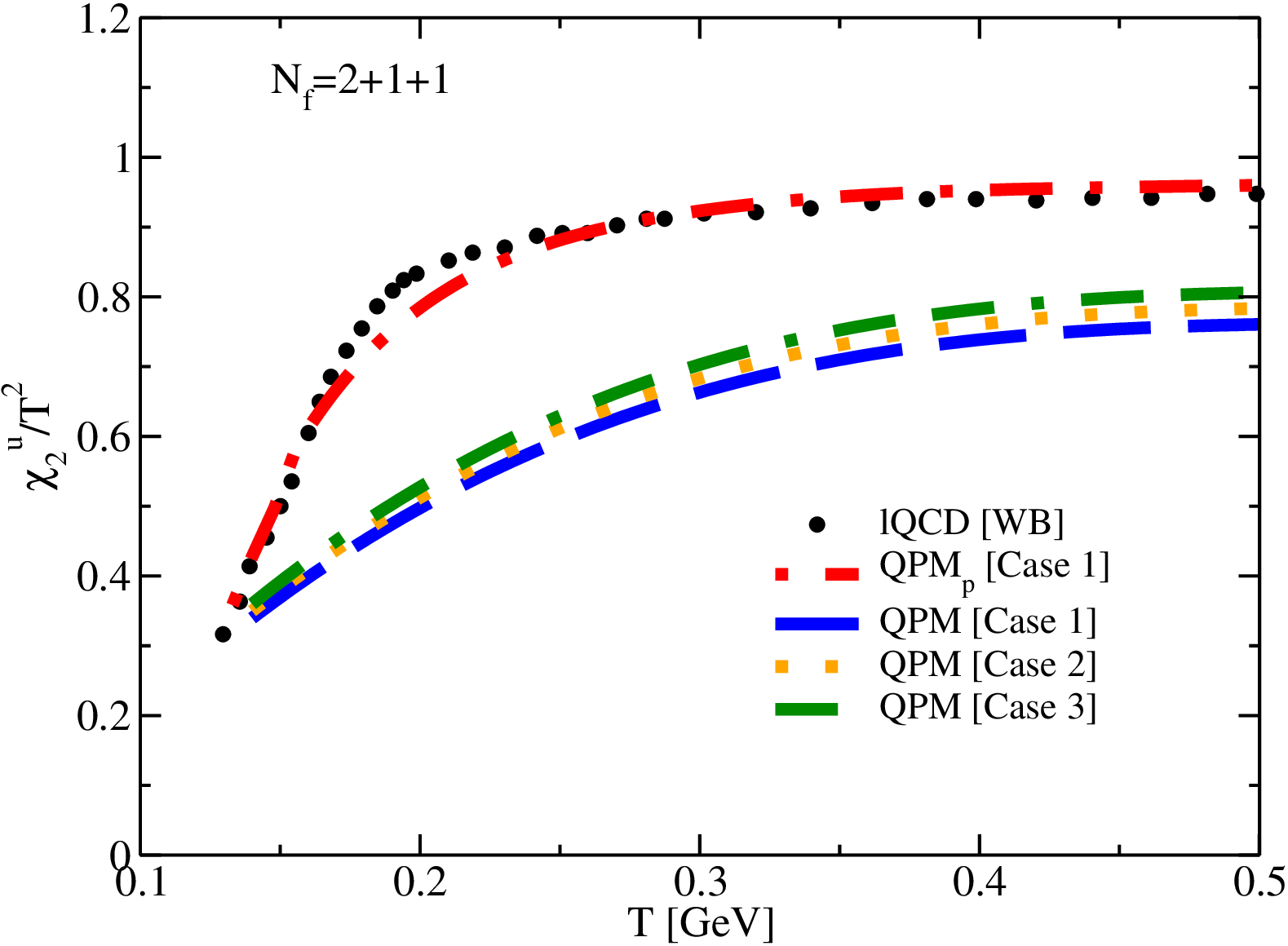}
\caption{Light quark number susceptibilities $\chi_u$ as functions of the temperature, for the different charm quark mass parametrizations, in the standard $QPM$ and for the extended $QPM$ model with momentum dependent quasi-particle masses.
The lQCD data have been taken from Ref.\cite{Bellwied:2015lba}.
}\label{fig:9}
\end{figure}
In Fig.s \ref{fig:8} and \ref{fig:9}, we show the strange $\chi_s/T^2$ and light $\chi_u/T^2$ susceptibilities in both the standard and extended $QPM$. 
As shown, the $QPM$ without momentum dependent quasi-particle masses underestimates the lQCD data quite significantly in the range of temperature explored, for all the different charm mass parametrizations. The main source of discrepancy is the large value of the quark thermal mass in the standard $QPM$, an aspect which was already noted in Ref.\cite{Plumari:2011mk}.
In the same plots, we also show the good agreement between the strange and light susceptibilities $\chi_s$ and $\chi_u$ in $QPM_{p}$ with respect to the lQCD data \cite{Bellwied:2015lba}, in agreement with the result found in Ref. \cite{Berrehrah:2015vhe, Berrehrah:2016vzw} for the light quark susceptibility.
This result is due to the momentum dependent quark mass which leads to a smaller 'thermal average mass' wrt the standard $QPM$, leading to an extra contribution in the susceptibility which permits the very good agreement with lQCD also in the vicinity of phase transition. 
\begin{figure}[ht]\centering
\includegraphics[width=1.0\linewidth]{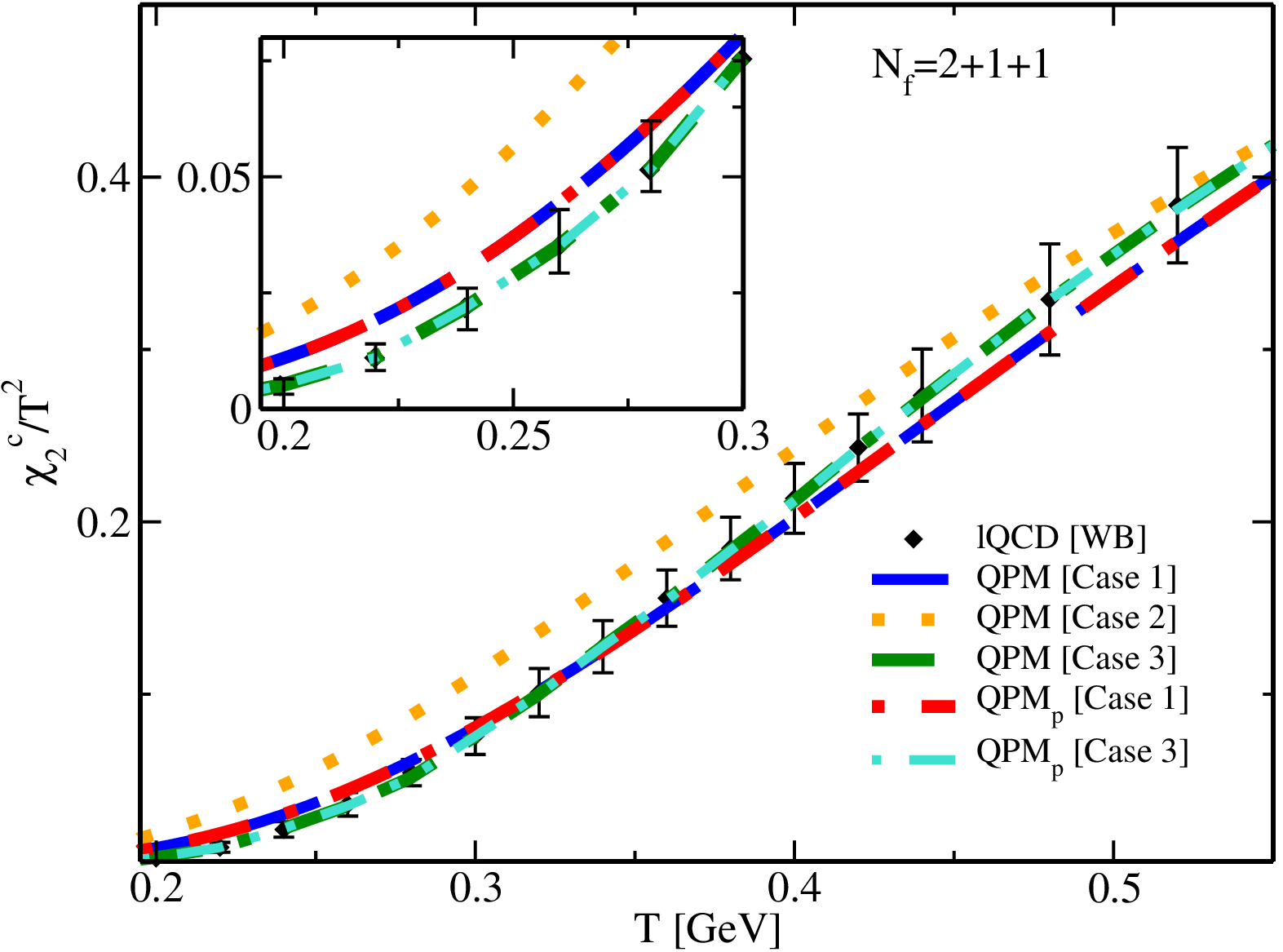}
\caption{Quark number susceptibilities $\chi_c$ as functions of the temperature. Different lines correspond to the different cases considered in the text. The black points are lQCD data taken from Ref.\cite{Bellwied:2015lba}. (Inset) Zoom-in the small temperature region.}\label{fig:10}
\end{figure}
Finally, in Fig.\ref{fig:10} we show the charm quark susceptibility $\chi_c/T^2$ for both standard and extended $QPM$. We compare the results of $\chi_c/T^2$ for the different cases considered in the previous Section. The aim is to analyze how different temperature parametrizations for the charm quark mass can affect the description of lQCD data for $\chi_c/T^2$. We recall that in these calculations, we have discarded the momentum dependence of charm quark mass, but we have checked that its impact on the quantities studied here is quite marginal. Our results show that the $\chi_c/T^2$ evaluated for a fixed charm quark mass $m_c=1.5 \, GeV$ (\textit{Case 1}) both for simple $QPM$ (blue dashed line) and $QPM_{p}$ (red double dot-dashed line) give the same temperature dependence of the susceptibility and a good agreement with the lQCD data. However, the last are underestimated by a small percentage only at high temperatures $T\gtsim 4 T_c$ while for  $T \approx 0.2-0.3$ $GeV$, the lQCD data are somewhat overestimated as shown in the inset of Fig. \ref{fig:10}.
Both these results are similar to the \textit{Case 3} where the charm mass was constrained by the charm number fluctuations, which obviously describe the charm quark susceptibility.
On the other hand, if we consider a thermal mass for the charm quark with the same parametrization of T dependence of light quarks $m_c^2=m_{c0}^2+\frac{N_c^2-1}{8 N_c}g^2 [T^2+\frac{\mu^2_c}{\pi^2}]$ with $m_{c0}=1.3$ $GeV$ (\textit{Case 2}), the thermal average of charm quark mass is smaller than $m_c=1.5 \, GeV$, leading to an overestimation of $\chi_c/T^2$ by a significant amount in the range of temperature considered. Even if a constant $m_c=1.5$ $GeV$ corresponds on average to a reasonable good $\chi_c(T)$ (at least up to $T\approx T_c$), the temperature dependence of charm quark susceptibility in the low T region seems to imply an increasing $m_c(T)$ from about $1.5$ $GeV$ at $T\approx 2 T_c$ up to $1.9$ $GeV$ at $T=T_c$. Furthermore, a $m_c\approx 1.3$ $GeV$ appears to be too light, leading to overestimate $\chi_c/T^2$ and also an $m_c(T)$ increasing with T is disfavored. Comparing the three different cases, we note that a $\frac{dm_c}{dT} < 0$ plays a relevant role and it is necessary to obtain the correct temperature evolution of $\chi_c(T)$ especially between $T_c$ or $2$ $T_c$. 

\section{Boltzmann transport equation and transport coefficients}

In this section we are interested to study both the HQs transport coefficients and the time evolution of the HQs phase-space distribution function. The main approach used to describe the evolution of HQs is the relativistic transport equation. Therefore, we remind the basic properties of the relativistic Boltzmann transport equation, applied to a system of HQs interacting with a bulk medium composed of light quarks and gluons.
We also briefly describe the derivation of HQ transport coefficients, i.e. drag and diffusion in momentum space, which are directly connected to the spatial diffusion coefficient $D_s$ of HQs. This coefficient is one of the most important parameters used to obtain information on HQ thermalization time and it can be also evaluated in lQCD. Therefore, our intent is to make a comparison between our results for $D_s$ obtained in the different $QPMs$ introduced in the previous Section and the lQCD calculations. 
Towards the end of this Section, we will also make a comparison between the $D_s$ in standard and in momentum dependent $QPM$, also extending the evaluation of the $D_s$ in the infinite charm quark mass limit, following the same idea of Ref.\cite{Sambataro:2023tlv}. This aspect is necessary to appropriately compare our results to lQCD calculations, which are evaluated in the infinite mass limit of HQs.\\

The Boltzmann transport equation can be expressed by the following integro-differential equation: 
\begin{eqnarray}\label{Eq:Boltzmann}
p^\mu\partial_\mu f_Q(x,p)=C[f_q,f_g,f_{Q}](x,p) 
\end{eqnarray}
where $C[f_q, f_g, f_Q](x,p)$ is the relativistic
Boltzmann-like collision integral describing the short range interaction between heavy quark and particles of the medium, with phase-space distribution functions $f_Q(x,p)$ 
 and $f_{q,g}(x,p)$ respectively. In the following discussion we will consider only two-body collisions and the collision integral $C[f_q, f_g, f_Q] (p)$ can be then expressed by the following relation:
\begin{align}\label{int_finale}
\begin{split}
&C[f]=\frac{1}{2E_p}\int\frac{d^3\textbf{q}}{2E_q (2\pi)^3}\int\frac{d^3\textbf{q}'}{2E_{q'}(2\pi)^3}\int\frac{d^3\textbf{p}'}{2E_{p'}(2\pi)^3}
\\&\cdot\frac{1}{d_Q}
\sum_{g,q,\bar q}|{\cal M}(g(q,\bar q), Q\rightarrow  g(q,\bar q), Q))|^2 
\\&\cdot (2\pi)^4\delta^4(p+q-p'-q')
[f_Q(\textbf{p}')\hat{f}(\textbf{q}')-f_Q(\textbf{p})\hat{f}(\textbf{q})]
\end{split}
\end{align}  
where \textbf{p} (\textbf{q}) and \textbf{p$\prime$} (\textbf{q$\prime$}) are the initial and final momenta of heavy quark (plasma particle) respectively and $|{\cal M_Q}|^2$ is the transition amplitude of the process. The collision integral of Eq.\eqref{int_finale} can be also expressed in relation to the rate of collisions $\omega(\textbf{p},\textbf{k})$ between HQ and bulk particles:
\begin{equation}\label{integrale_coll}
C[f]=\int d^3\textbf{k} [\omega (\textbf{p}+\textbf{k},\textbf{k})f(\textbf{p}+\textbf{k})-\omega(\textbf{p},\textbf{k})f(\textbf{p})].
\end{equation}
where the rate of collision $\omega(\textbf{p},\textbf{k})$ is given by:
\begin{equation}\label{rate}
\omega(\textbf{p},\textbf{k})=d_{QGP} \int \frac{d^3\textbf{q}}{(2\pi)^3} \hat f(\textbf{q})v_{q,p}
\frac{d\sigma_{p,q\rightarrow p-k,q+k}}{d\Omega}.
\end{equation}

with \textbf{k} the transferred momentum in the collision and $d_{QGP}$ the number of degrees of freedom of the particle in collision with the heavy quark. The collision rate is also expressed in relation to the relative velocity $v_{q,p}$ between particles in the collision, whereas the differential cross section of the scattering process is given by:
\begin{equation}\begin{split}
\frac{d\sigma_{p,q\rightarrow p-k,q+k}}{d\Omega}=\frac{1}{(2\pi)^6}\frac{1}{v_{p,q}}\frac{1}{2E_q}\frac{1}{2E_p
}\frac{1}{d_Q d_{QGP}}
\sum|{\cal M_Q}|^2\\\times\frac{1}{2E_{q+k}}\frac{1}{2E_{p-k}}(2\pi)^4\delta(E_p+E_q-E_{p-k}-E_{q+k})
\end{split}
\end{equation}

In order to simplify the calculation of the non-linear integro-differential Boltzmann equation, the \textit{Landau approximation} is often employed leading to a relativistic Fokker-Planck equation. Therefore, assuming that during collision the transferred momentum $\textbf{k}$ is small, we can operate an expansion of the integrand:
\begin{align}
\begin{split}
f(\textbf{p}+\textbf{k})\omega(\textbf{p}+\textbf{k},\textbf{k})=&f(\textbf{p})\omega(\textbf{p},\textbf{k})+\textbf{k}\cdot\frac{\partial}{\partial \textbf{p}}(\omega f)\\&+\frac{1}{2}k_ik_j\frac{\partial^2}{\partial p_i \partial p_j}(\omega f)+...
\end{split}
\end{align} 
Defining the following quantities:
\begin{equation}\begin{split}\label{coeff}
A_i(\textbf{p},T)&=\int d^3kk_i\omega(\textbf{p},\textbf{k})\\
B_{i,j}(\textbf{p},T)&=\frac{1}{2}\int d^3kk_ik_j\omega(\textbf{p},\textbf{k})
\end{split}
\end{equation}
the collision integral $C[f]$ in Eq.\eqref{int_finale} becomes:
\begin{eqnarray}\label{fokker-planck}
\frac{df(\textbf{p})}{dt}&=&\frac{\partial}{\partial p_i}\left[A_i(\textbf{p},T)f(\textbf{p})+\frac{\partial}{\partial p_j}[B_{i,j}(\textbf{p},T)f(\textbf{p})]\right]. \nonumber \\
\end{eqnarray}
the above equation is the \textit{Fokker-Planck equation}. The quantities defined by the Eq.\ref{coeff} are the drag and diffusion coefficients that govern the propagation of HQs in the thermal bath at temperature $T$. If we consider an isotropic medium, we can express the drag and diffusion coefficients by the following relations:
\begin{equation}\begin{split}
A_i(\textbf{p},T)&=A(p,T)p_i.\\
B_ {i,j}(\textbf{p},T)&=B_L(p,T)P_{i,j}^{||}(\textbf{p})+B_T(p,T)P_{i,j}^{\perp}(\textbf{p}).
\end{split}
\end{equation}
where $P_{i,j}^{||}(\textbf{p})=p_ip_j/\textbf{p}^2$ and $P_{i,j}^{\perp}(\textbf{p})=\delta_{i,j}-(p_ip_j/\textbf{p}^2)$ are the projection operators on the longitudinal and transverse momentum components so that the diffusion coefficient is expressed by a longitudinal $B_L$ and a transversal $B_T$ component with respect to the HQ momentum.
Using the definition in Eq.\eqref{coeff}, we get the following expression for $A_i$:
\begin{eqnarray}\label{drag}
  & & A_i(\textbf{p},T)= \nonumber \\
  &=&\frac{1}{2E_p}\int\frac{d^3\textbf{q}}{2E_q (2\pi)^3}\int\frac{d^3\textbf{q}'}{2E_{q'}(2\pi)^3}\int\frac{d^3\textbf{p}'}{2E_{p'}(2\pi)^3}\nonumber \\
  &\times&\frac{1}{d_Q}\times \sum|{\cal M_Q}|^2(2\pi)^4\delta^4(p+q-p'-q') \nonumber \\
  &\times&\hat f(q)[(p-p')_i]\equiv \left\langle \left \langle(p-p')_i\right \rangle \right\rangle.
\end{eqnarray}
Finally, the drag coefficient can be calculated as follows:
\begin{equation}\begin{split}\label{drag_comp}
A(p,T)&=p_iA_i/p^2=\\&=\left\langle \left \langle\textbf{1}\right \rangle \right \rangle-\left\langle \left \langle\textbf{p}'\cdot\textbf{p}\right \rangle \right \rangle/p^2.
\end{split}
\end{equation}

\begin{figure}[ht]\centering
\includegraphics[width=1.0\linewidth]{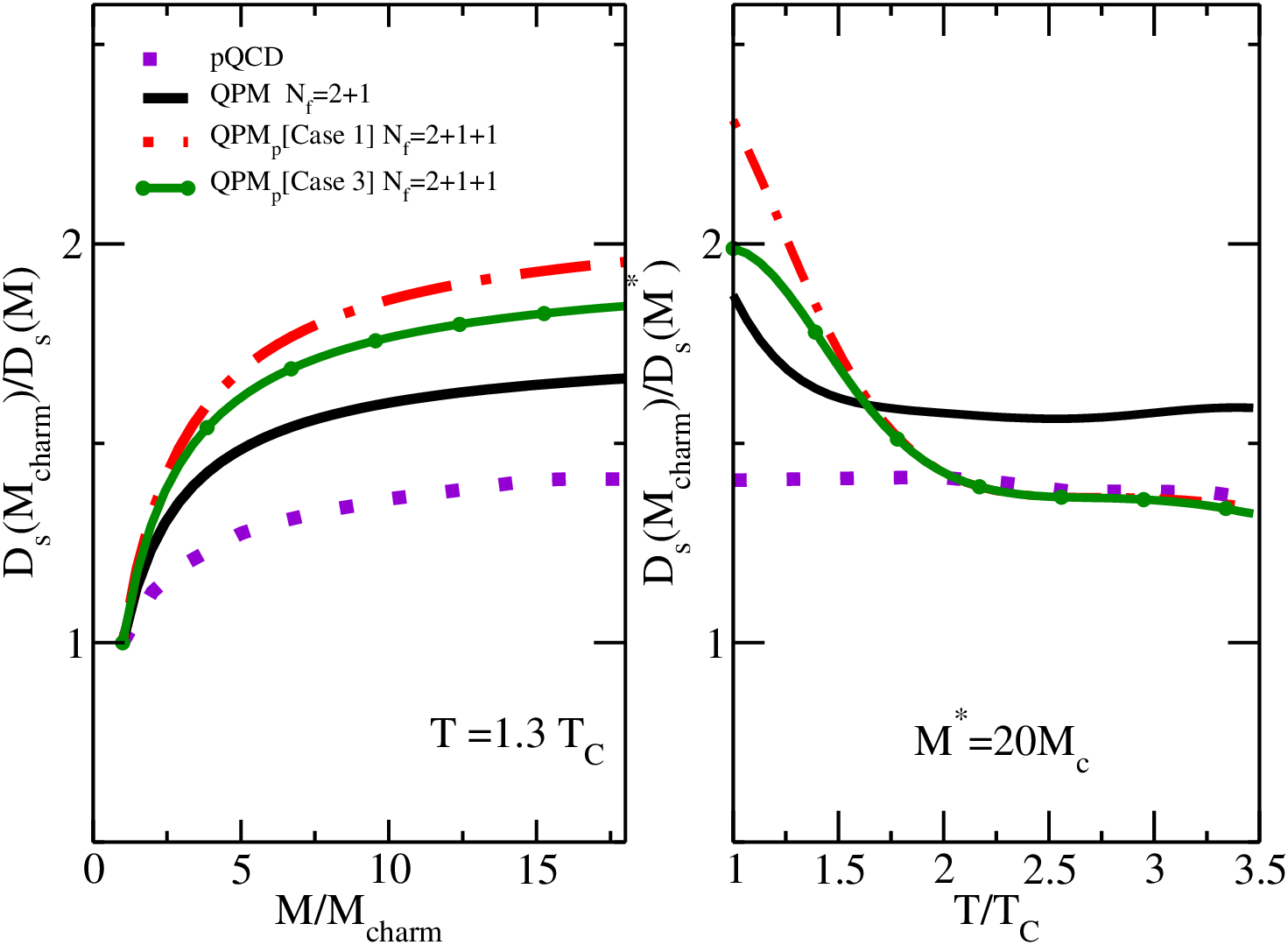}
\caption{(left) Ratio $D_s(M_{charm})/D_s(M)$ as function of $M/M_{charm}$ for $QPM$ (black solid line), $QPM_p$ with Case 1 (red dot-dashed line) and Case 3 (dark green line) charm mass parametrizations and also within pQCD (dashed purple line) at $T=200 MeV\approx 1.3T_c$. (right) Ratio between $D_s(charm)$ and $D_s(M\rightarrow \infty)$ as a function of the scaled temperature $T/T_c$ for all the previous cases.}\label{fig:12}
\end{figure}

\begin{figure}[ht]\centering
\includegraphics[width=1.0\linewidth]{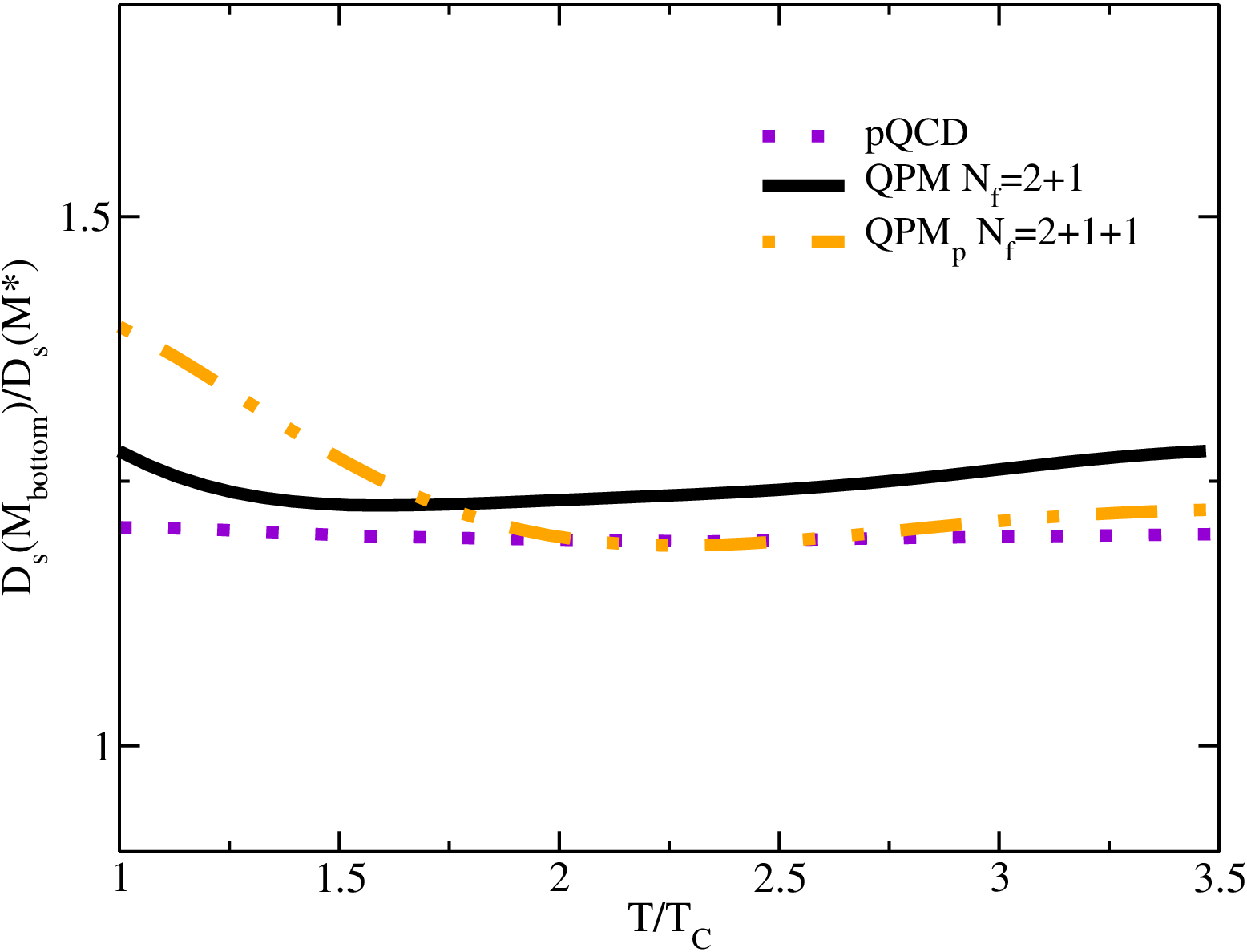}
\caption{Ratio $D_s(M_{bottom})$/$D_s(M\rightarrow \infty)$ as a function of the scaled temperature $T/T_c$ for $QPM$ $N_f=2+1$ and $QPM_p$ $N_f=2+1+1$ for a constant mass of bottom quark $M_b=4.7$ $GeV$}\label{fig:13}
\end{figure}

In the static limit $p\rightarrow 0$, we have that $B_T=B_L=D_p$ and we write $A=\gamma$.
More details about the standard approach to evaluate the integral in Eq.\eqref{drag} can be found in Ref.\cite{Svetitsky:1987gq, Berrehrah:2014kba, Sambataro:2020pge}.\\
The diffusion in momentum space can
be shown to lead to diffusion in position: one of the most significant parameters to describe the HQ interaction with the medium is the \textit{spatial diffusion coefficient} $D_s$, which can be expressed by:
\begin{equation}\label{Ds}
D_s=\frac{T^2}{D_p}=\frac{T}{M_{HQ}\gamma}
\end{equation}
where we have imposed the Einstein relation $D_p=TM\gamma$ to reduce from the first to the second equation.
In our calculation we use the second relation which is equivalent to the first one because the static limit $(p \to 0)$ is not significantly affected by the violation of fluctuation-dissipation theorem, as instead seen in the finite momentum region \cite{Sambataro:2020pge}. In particular, the validity of the Einstein relation is strengthened when we consider the infinite mass limit of HQs which we take into account in order to appropriately compare to the lQCD calculations for $D_s$.\\
\begin{figure}[t!]
\centering
\includegraphics[width=1.0\linewidth]{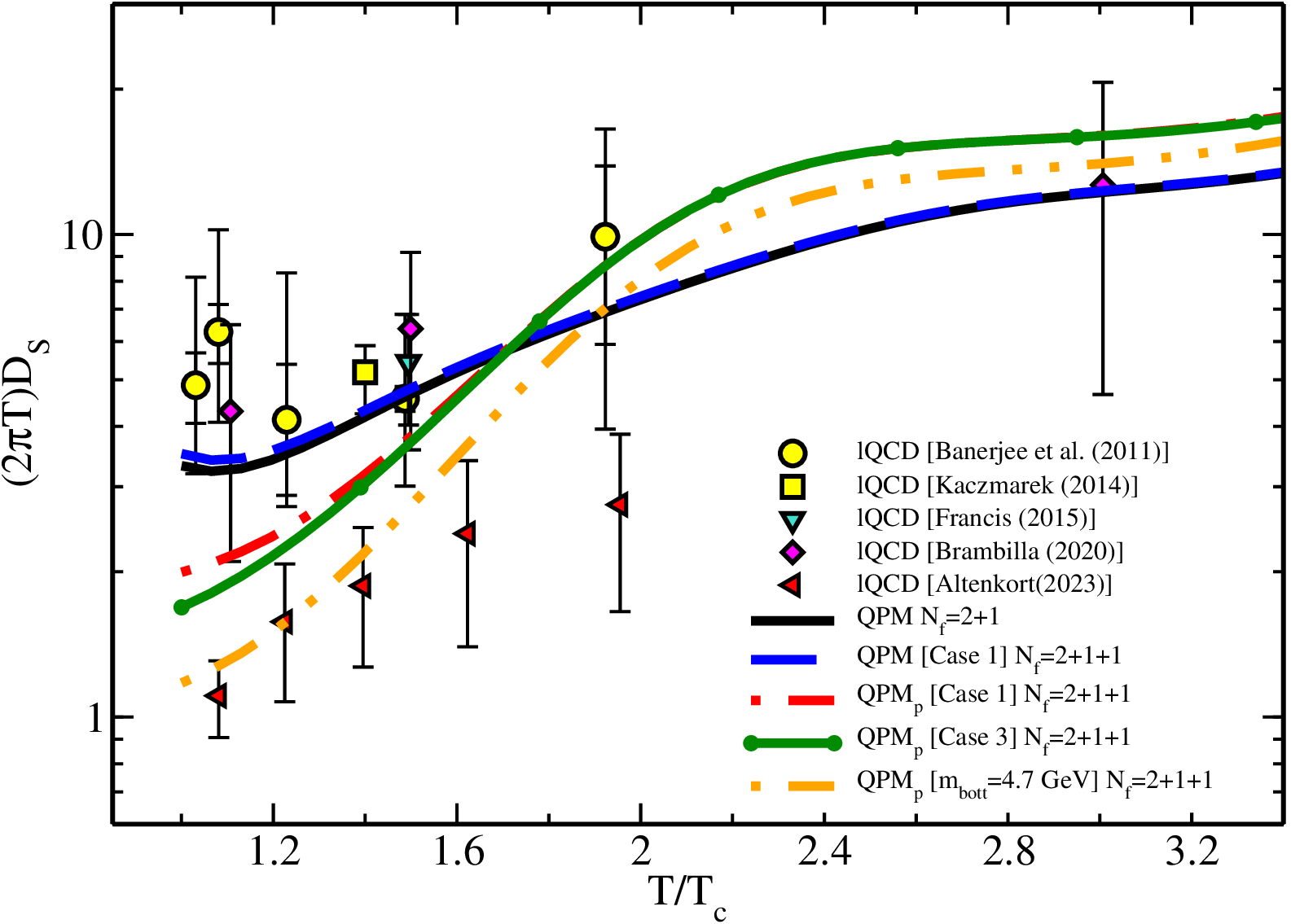}
\caption{Spatial diffusion coefficient $D_s$ as a function of the scaled temperature $T/T_c$. 
Different curves are for different $QPMs$ discussed in the text: the standard $QPM$ with $N_f=2+1$ (black solid line), the standard $QPM$ with $N_f=2+1+1$ (blue dashed line) and the extended approach including the momentum dependent quasi-particle masses $QPM_{p}$ (red dot-dashed line). The dark green curve stands for the extended $QPM$ for the Case 3. The points are the available lQCD data, taken from \cite{Banerjee:2011ra, Kaczmarek:2014jga, Francis:2015daa, Brambilla:2020siz, Altenkort:2023oms}.}\label{fig:11}
\end{figure}
Furthermore, the scattering matrices $|\cal M_Q|$$^2$ are evaluated considering the tree level diagrams of the relevant processes, with the effective coupling $g(T)$ which is extracted from the different $QPM$ cases discussed in previous section. The
details of this procedure, which follows the same strategy of the Dynamical Quasi-Particle Model (DQPM), can be found in Ref.s \cite{Berrehrah:2013mua, Sambataro:2020pge, Sambataro:2022xzx, Sambataro:2023tlv}. In Fig. \ref{fig:11}, we show the spatial diffusion coefficient $D_s$ as function of scaled temperature $T/T_c$. We compare the different $QPMs$ discussed in the previous section: the standard $QPM$ with $N_f=2+1$ flavours (black solid line), the standard $QPM$ with $N_f=2+1+1$ including charm with mass $m_c=1.5$ $GeV$ [Case 1](blue dashed line) and the $QPM_{p}$ extended approach with momentum dependence for constant charm mass $m_c=1.5$ $GeV$ [Case 1] (red dot-dashed line).  In addition, we also evaluate the $D_s$ for the $QPM_p$ in the case the in-medium charm quark mass is constrained by quark number fluctuations and assumes a temperature dependence as of Case 3 (green solid curve) and for the bottom quark with constant mass $m_{bott}=4.7$ $GeV$ (dot-dashed orange line). As shown by comparing the blue dashed line with the black solid line, the $D_s$ evaluated from the $QPM$ including charm quarks does not show significant difference wrt the standard $QPM$ with $N_f=2+1$ flavours. On the other hand, we see that for $T\lesssim T_c$ the $D_s$ evaluated in the extended model $QPM_{p}$ exhibits a strong decrease in the low temperature region $T < T_c$, while it grows toward the pQCD estimate faster than the standard approach at higher temperature. This behaviour can be explained in relation to the effective coupling of $QPM_p$ shown in Fig.\ref{fig:4}, which is higher than the standard one at low temperatures and smaller in the region of high temperatures. 
Moreover, at higher temperature the momentum dependence of the quasi particle masses in $QPM_{p}$ has the effect to recover the pQCD limit, resulting in an enhancement of the spatial diffusion coefficient $D_s$.
Our results show that within the $QPM_{p}$ the $D_s$ gets closer to the new lQCD data points: these are the more pertinent ones to compare our data to, since they include dynamical fermions differently from the calculations until 2020, which are instead evaluated for a \textit{quenched} medium. Furthermore, the case of $QPM_{p}$ with $m_c(T)$ shows a greater reduction of $D_s$ at small temperature wrt the constant mass case, due to the effective larger charm quark mass in the same temperature region. We also notice that the $D_s(T)$ for the bottom quark with constant mass $m_{bott}=4.7$ $GeV$ is already very close to the lQCD data points especially in the low temperature region.\\
In order to have a consistent comparison between our results and lQCD data, we also remind that the lQCD calculations reported here are evaluated in the infinite mass limit for HQs, while in $QPM$ this limit is not yet reached at the charm mass scale, an aspect that was studied and shown in Ref. \cite{Sambataro:2023tlv}. In particular, within kinetic theory in the $M/T \rightarrow \infty$ limit, we expect the drag coefficient to scale linearly with $M_{HQ}$, leading to a mass independent $D_s$ which provides a general measure of the QCD interaction. However, in $QPM$ approach the
$D_s$ is strongly mass-dependent around the charm mass scale, reaching a saturation value only for masses $M_Q \sim 10 M_{charm} \gtsim 
15$ $GeV$, giving $D_s(M_{charm})/D_s(M \rightarrow \infty) \simeq 1.9$ at $T=T_c$. This result suggests that at the charm mass scale the infinite mass limit used in lQCD is not yet reached.\\ 
In this context, we have studied how the $D_s$ mass dependence changes in the $QPM_p$ in comparison to the standard $QPM$. In Fig. \ref{fig:12} we plot, in the left panel, the ratio between $D_s(M_{charm})$ and $D_s(M)$ as function of $M/M_{charm}$ at $T=200$ $MeV$ $\approx 1.3$ $T_c$ and, in the right panel, the ratio $D_s(M_{charm})/D_s(M^{\ast})$ as function of scaled temperature $T/T_c$. In this calculation, $M^\ast$ is the mass of a fictitious super-heavy quark, which we set as the infinite mass limit. In particular, we have chosen a large value $M^\ast=20M_c$ in order to completely reach the $D_s$ saturation value. 
In proximity to the critical value of temperature $T_c$, the $D_s$ for charm quark in $QPM_p$ is quite larger than the one for $M_{HQ}\rightarrow \infty$, even larger than the one in $QPM$, but at $T\approx 2$ $T_c$ the difference reduces to about a $30$ $\%$ similar to what is obtained in a leading-order pQCD framework. Furthermore, the discrepancy between the bottom mass scale in $QPM_p$ and the infinite mass limit at $T=T_C$ is smaller of about a factor $1.5$ wrt the charm quark as can be seen in Fig.\ref{fig:13}. We notice that, when we include a $m_c(T)$ following the Case 3, the $QPM_p$ is closer to the infinite mass limit at $T_c$ of about $15 \%$ wrt the case with constant mass, while the two sets of data are similar for $T>2T_c$. This result suggests that the temperature dependence of charm quark of Case 3 seems to be the most reasonable in order to better describe the lattice QCD $D_s(T)$, especially in the small T region.\\
The most relevant result is shown in Fig.\ref{fig:14}, where we compare the temperature dependence of the $D_s(T)$ in the standard $QPM$ and in the extended $QPM_{p}$ in the infinite mass limit for HQs with lQCD data. Our results show that the $QPM_p$ is in a quite better agreement to the new lQCD data by  Ref.\cite{Altenkort:2023oms} wrt the standard $QPM$. We also notice that even at finite charm and bottom mass, $QPM_p$ predicts a $D_s(T)$ quite closer to lQCD wrt $QPM$. This is particularly valuable considering that $QPM_{p}$ is an approach able to correctly describe the lQCD thermodynamics ($\epsilon(T), P(T), s(T)$) as well as the light quark $\chi_q(T)$ and the charm quark $\chi_c(T)$ susceptibilities. Therefore, it represents in a realistic simulation a significant improved model to employ for the prediction of experimental observables $R_{AA}(p_T)$ and $v_n(p_T)$.
\begin{figure}[ht]\centering
\includegraphics[width=1.0\linewidth]{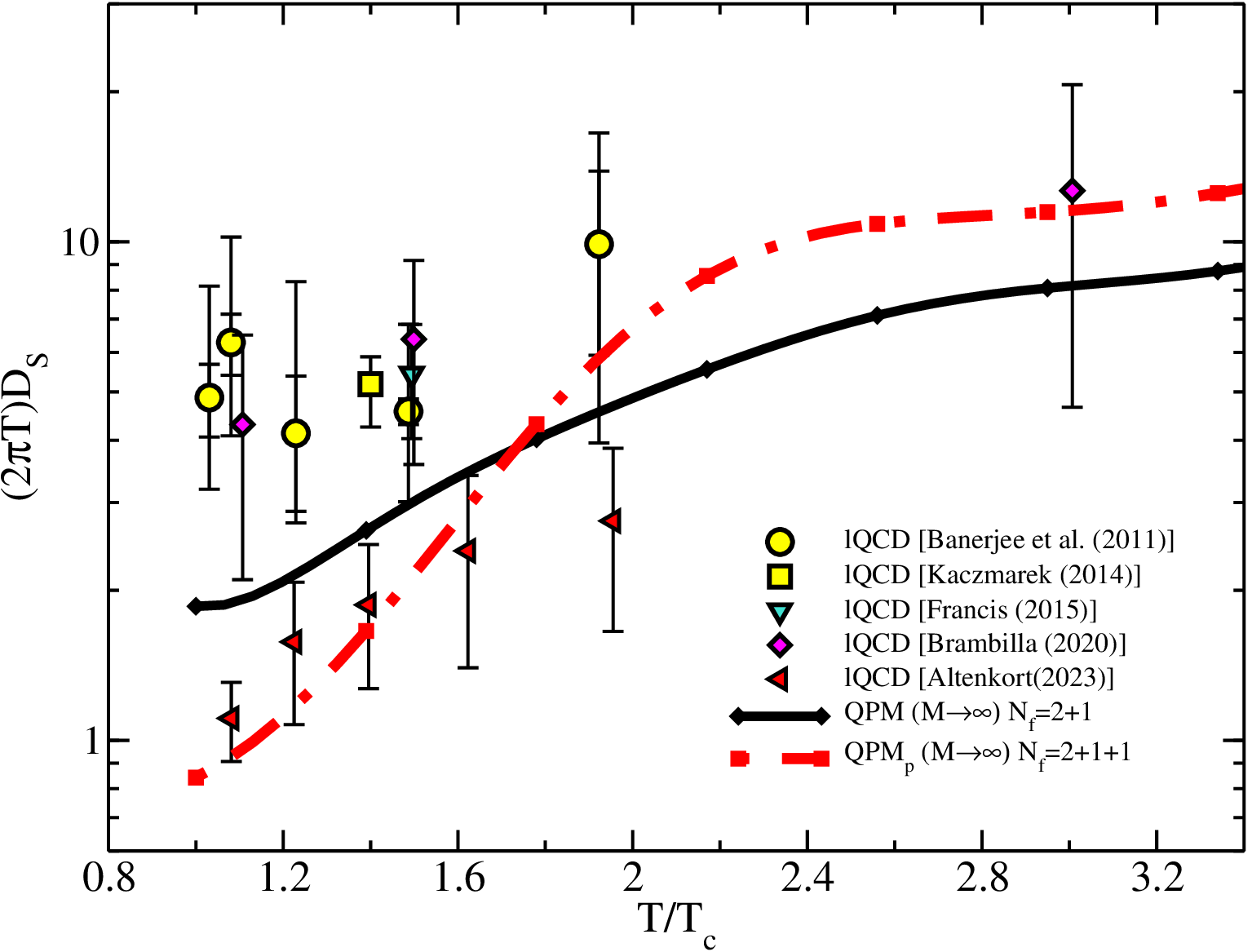}
\caption{Spatial diffusion coefficient $D_s$ as a function of the scaled temperature $T/T_c$. Black solid line refers to the standard $QPM$ in the infinite mass limit and red dot-dashed line to the extended case $QPM_{p}$ in the same limit of $QPM$. Points refer to lQCD data with the same legend as in Fig.\ref{fig:11}.}\label{fig:14}
\end{figure}
\section{Heavy Quarks momentum evolution in the QGP}

\begin{figure}[ht]\centering
\includegraphics[width=1.0\linewidth]{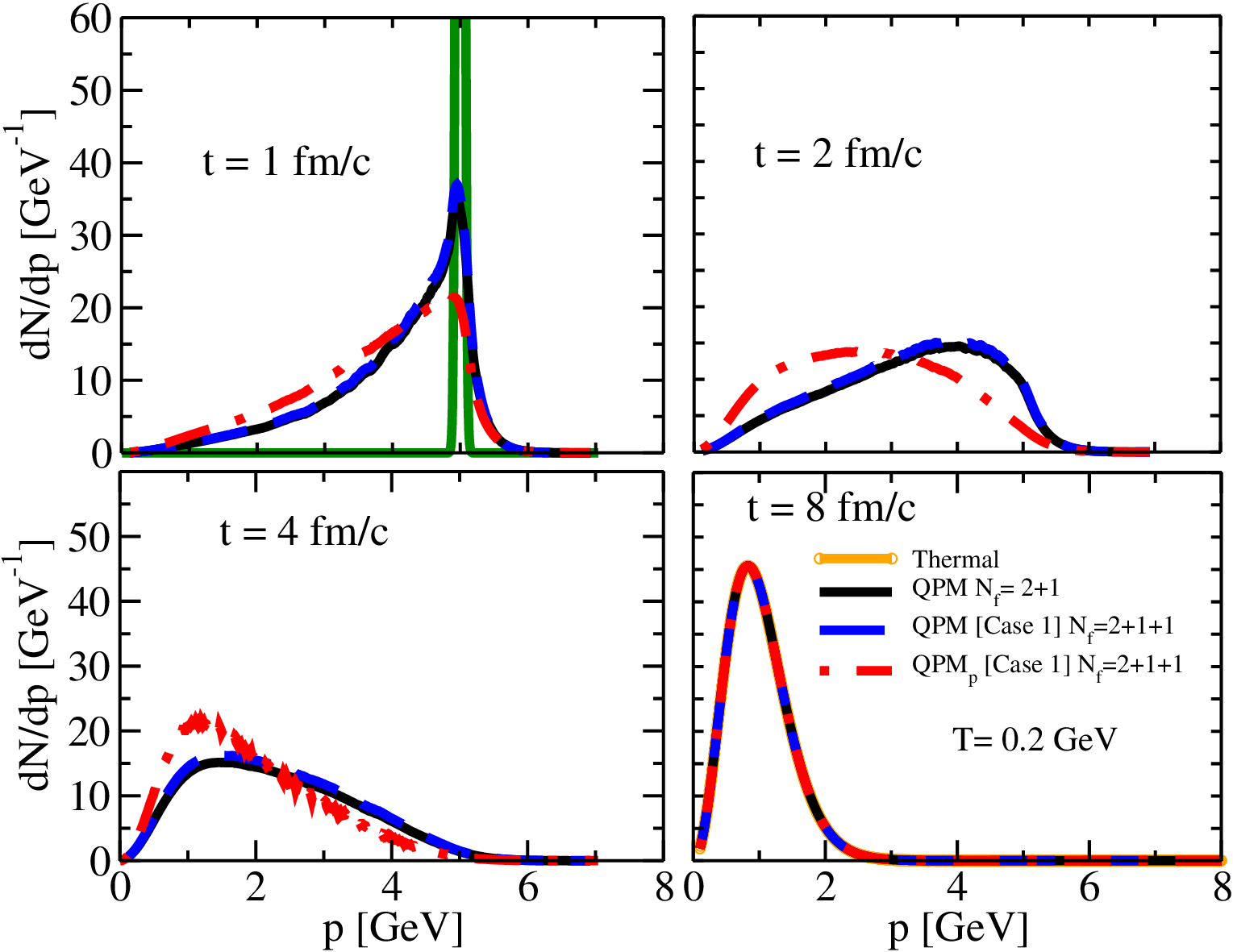}
\caption{
Charm quark momentum distribution as a
function of the charm quark momentum at fixed temperature $T=0.2 \, GeV$ and four different times $t = 1 fm/c$ (left upper panel) and $t = 2 fm/c$ (right upper panel), $t = 4 fm/c$ (left lower panel) and $t = 8 fm/c$ (right lower panel). Black solid lines stand for the $QPM$ with $N_f=2+1$, blue dashed lines for the $QPM$ with $N_f=2+1+1$ [Case 1] and red dot-dashed lines for the case with momentum dependence $QPM_{p}$ [Case 1].
}\label{fig:15}
\end{figure}

\begin{figure}[ht]\centering
\includegraphics[width=1.0\linewidth]{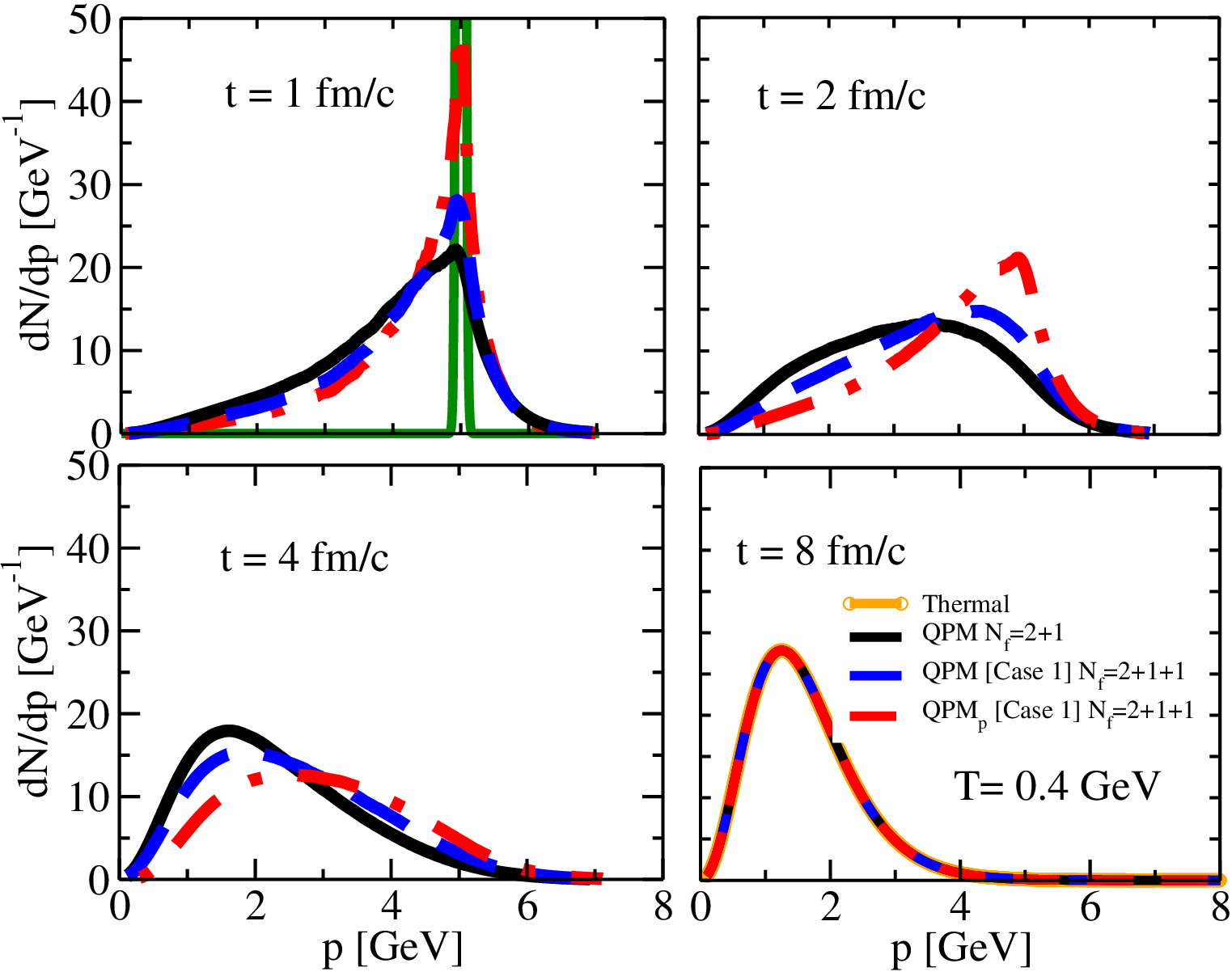}
\caption{
Charm quark momentum distribution as a
function of the charm quark momentum at fixed temperature $T=0.4 \, GeV$ and four different times $t = 1 fm/c$ (left upper panel) and $t = 2 fm/c$ (right upper panel), $t = 4 fm/c$ (left lower panel) and $t = 8 fm/c$ (right lower panel). Same legend as in Fig. \ref{fig:15}.
}\label{fig:16}
\end{figure}

In this section, we discuss the time evolution of HQs within Boltzmann transport equation for the different $QPM$ and $QPM_p$ cases considered. We are interested in the evolution of the HQ distribution function $f_{Q}(x,p)$ in a thermal bulk of light quarks and gluons in equilibrium in a box with constant temperature $T$.\\
Assuming that the plasma
is uniform, i.e. HQs distribution function is $x$ independent, we can discard the field gradients in the Boltzmann equation of Eq.\eqref{Eq:Boltzmann} and it becomes:
\begin{equation}\label{Eq:Boltzmann_discr}
\frac{\partial f_{Q}}{\partial t}=\frac{1}{E_{Q}}C[f_q,f_g,f_{Q}].
\end{equation}
Numerically, it can be solved by the following time discretization:
\begin{equation}\label{Eq:discr}
f(t+\Delta t,p)=f(t,p)+\frac{\Delta t}{E_{Q}}C[f] + O(\Delta t^2).
\end{equation}
The numerical solution of Boltzmann equation is obtained by a code in which we have discretized the time and the HQ momentum \textit{p}. Furthermore, we have implemented a Monte-Carlo integration method for the full collision kernel, as discussed in Ref.\cite{Sambataro:2020pge}. Different tests have been performed in order to check the convergence of the solution. In order to study the difference in the evolution obtained for the different $QPMs$, we show the resulting phase-space distribution function of charm quarks evolving in the plasma of quarks and gluons for two different values of temperature, $T=0.2 \, GeV$ and $T=0.4 \, GeV$. In the following calculations, the initial charm quark distribution is assumed to be approximately a delta distribution in momentum space centered at $p_0 = 5 \,GeV$, as shown by the green lines in the top left panels of both Fig.s \ref{fig:15} and \ref{fig:16}. 
Different lines represent different $QPM$ calculations: the standard $QPM$ with $N_f=2+1$ (black solid line), the one with $N_f=2+1+1$ including charm quarks with mass $m_c=1.5$ $GeV$ [Case 1] (blue dashed line) and the extension with momentum dependent quasi-particle masses $QPM_{p}$ with $m_c=1.5$ $GeV$ [Case 1] (red dot-dashed line). 
As shown by comparing red and blue lines, the $QPM_{p}$ has a slower dynamics at $T=0.4$ $GeV$ than the standard $QPM$ and faster dynamics at $T=0.2$ $GeV$. This behaviour is due to the fact that in the high temperature regime the $QPM_{p}$ is closer to a pQCD like dynamics with a larger $D_s$ than $QPM$, while it is characterized by a larger cross-section and smaller $D_s$ at lower temperatures, due to a stronger increase of the coupling $g(T)$. 
At $t > 4 \, fm/c $, as expected, in both cases the momentum distribution tends towards a thermal distribution as shown by the open square orange points in the right lower panels of Fig.s\ref{fig:15} and \ref{fig:16}. For the phenomenology of $HIC$, it is more relevant to see what is the global difference between $QPM$ and $QPM_p$ in an expanding and cooling system resembling the $HIC$ evolution. We discuss this aspect in the next subsection.

\subsection{Nuclear modification factor $R_{AA}$ in Boltzmann and extended $QPM$}

\begin{figure}[ht]\centering
\includegraphics[width=1.0\linewidth]{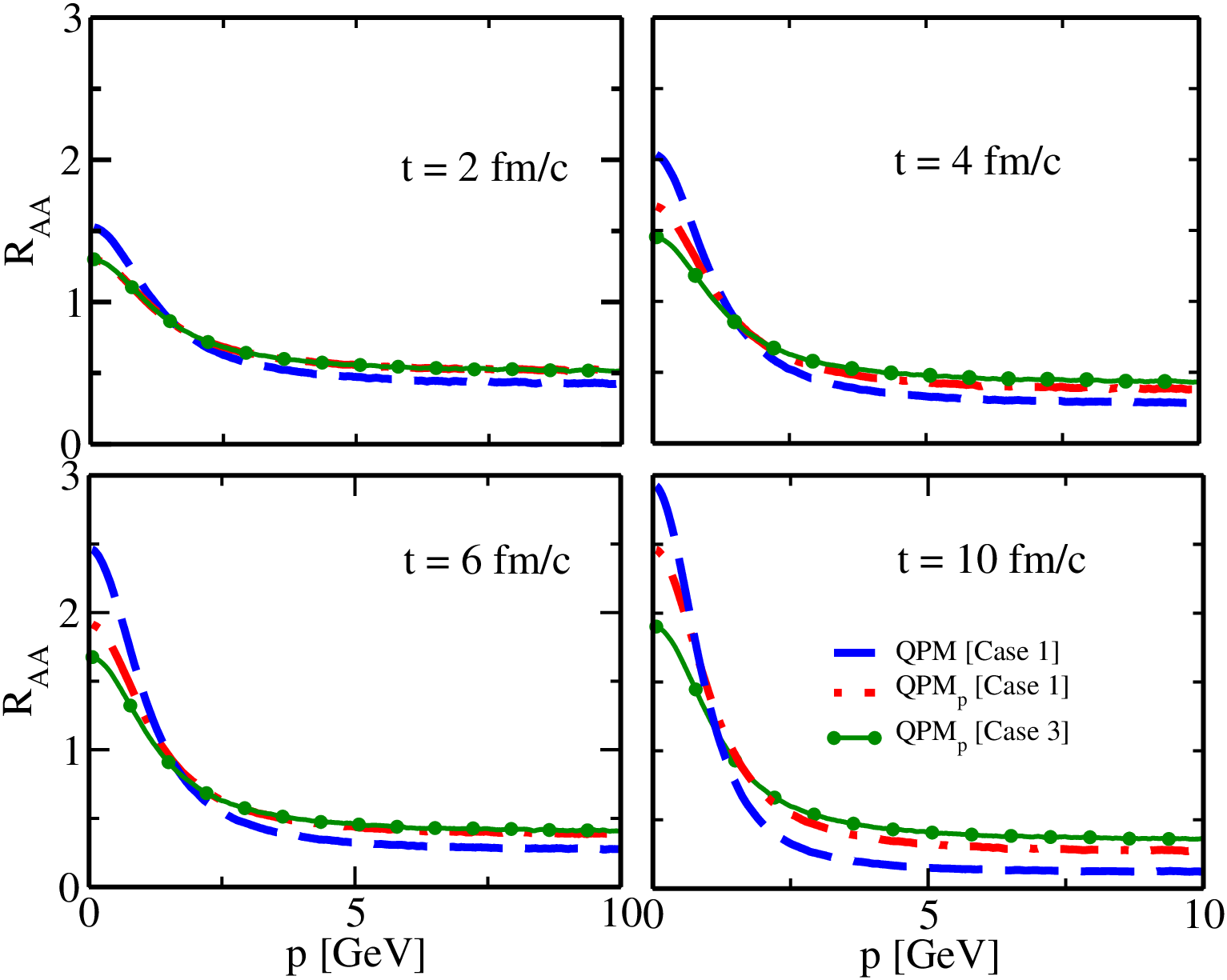}
\caption{$R_{AA}(p)$ as a function of charm momentum at different time steps $t=2,4,6,10$ $fm/c$. Different lines refer to: (blue dashed lines) $QPM$ with $m_c=1.5$ $GeV$ [Case 1], (red dot-dashed lines) $QPM_{p}$ with $m_c=1.5$ $GeV$ [Case 1] and (dark green curve) $QPM_p$ with temperature dependent charm quark mass [Case 3]. All cases are for $N_f=2+1+1$.}\label{fig:17}
\end{figure}

The \textit{nuclear modification factor} $R_{AA}$ represents one of the main observables investigated at both RHIC
and LHC energies in the heavy-flavour sector. In general, the $R_{AA}$ gives a quantitative estimate of the HQ-medium interaction, being expressed as the effective energy loss in AA collisions wrt the production in pp collisions. In this section, we show the impact of the $QPM_p$ with momentum dependent masses on the evolution of spectra in terms of $R_{AA}(p)$ for charm quarks. In these calculations, we simulate a $1+1D$ system with a temperature $T(t)$ that evolves with time as in 1D ideal hydrodynamics, i.e. described by the following behaviour:
$$T=T_0 (t / t_0)^{-\frac{1}{3}},$$
where $T_0 =0.55 \, GeV$ and $t_0 =0.3 \,fm/c$ are the typical initial temperature and time at LHC, respectively. Starting from the charm quark production in fixed order + next to-leading log (FONLL) which describes the D-meson spectra in pp collisions after fragmentation that we use as initial momentum distribution of charm quark $f_c(p,t_0)$, we evaluate the nuclear modification factor as $R_{AA}=f_c(p,t_f)/f_c(p,t_0)$, where $f_c(p,t_f)$ is the phase-space distribution function after Boltzmann evolution. In Fig.\ref{fig:17}, we show the nuclear modification factor $R_{AA}$ as a function of the charm quark momentum $p$ for both $N_f=2+1+1$ standard $QPM$ and extended $QPM_{p}$ with $m_c=1.5$ $GeV$ at different time steps ($t = 2, 4, 6, 10 \, fm/c$). Furthermore, we also evaluate $R_{AA}$ for the extended approach with $m_c(T)$ as of Case 3. Our results show that globally over the typical temperature evolution of uRHICs, the net effect is a reduction of $R_{AA}(p)$ at low $p$ in the $QPM_p$ wrt the one in simple $QPM$. The result is non trivial because as can be seen in Fig.\ref{fig:11}, the $D_s(T)$ of $QPM_p$ is larger than $QPM$ at $T > 1.7$ $T_c$ but smaller at lower temperature.
This effect is more significant for $QPM_p$ with charm mass from Case 3 wrt the Case 1 with $m_c=1.5$ $GeV$, in which a further reduction of $R_{AA}$ at low $p$ of about $15 \%$ is obtained even if its $D_s(T)$ is even lower but likely the large $m_c$ makes the global quenching smaller.
We stress that this result, for a realistic simulation, could give better agreement with the most recent available experimental data in the low $p$ region ($p<2$ $GeV$) of D meson $R_{AA}(p_T)$ which, in particular, is overestimated by our previous calculations within the standard $QPM$ even when the shadowing (discarded here) and the meson and baryon hadronization are included \cite{Plumari:2017ntm, Minissale:2020bif, Minissale:2023dct}.\\

\section{Conclusions}

In this paper, we have studied the extension of our Quasi-Particle Model ($QPM$) to a more realistic approach which includes momentum dependence in the quark and gluon masses ($QPMp$). The aim of this extension is to incorporate the dynamics driving the quasi-particle masses for $p\rightarrow \infty$ ($p >> \Lambda_{QCD}$) to the current quark masses, as also seen in Dyson-Schwinger or T-matrix approaches, an aspect that the simple $QPM$ with temperature-only dependent masses does not take into account. In this extended approach, named $QPM_p$, we correctly reproduce not only the EoS but also both light and strange quark susceptibilities. This result is due to the momentum dependent factors in light and strange particle masses that lead to an additional factor in the quark number susceptibility wrt to the standard $QPM$ which instead is known to underestimates the lQCD data. Furthermore, by upgrading $QPM$ with the recent lQCD calculations which include the charm quark ($N_f=2+1+1$), we have also studied the impact of different charm quark mass parametrizations as function of temperature on the charm susceptibility $\chi_c$: our results shows that a constant charm quark mass $m_c=1.5$ $GeV$ gives a reasonably good agreement with the available lQCD data for charm quark susceptibility. 
Instead a temperature dependent charm mass given by the same parametrized form as for the strange quark in $QPM$, i.e. $m_{c}^2= m_{c0}^2+\frac{N_c^2-1}{8N_c}g^2[T^2+\frac{\mu_c^2}{\pi^2}]$ with $m_{c0}=1.3$ $GeV$, does not provide a good description of charm quark lQCD susceptibility $\chi_c$, which is significantly overestimated in all the temperature region considered $T_c<T<4T_c$. 
One of the main results of this paper concerns the modification in the extended $QPM_p$ model of the spatial diffusion coefficient $2\pi T D_s$ as function of the temperature, also evaluated in the infinite mass limit for heavy quark in order to better compare to lQCD calculations. We have seen that the $QPM_p$ certainly provides a better quantitative agreement with new lQCD data which include dynamical fermions, wrt the previous $QPM$ modeling. This result is mainly due to the fact that the enhancement of the $g(T)$ in $QPM_p$ wrt simple $QPM$ for $T<2 T_c$ leads to stronger non-perturbative behaviour in the $D_s$ in the same temperature region. This is a remarkable result, since the $QPM_p$ correctly describes also both the lQCD EoS, light and strange quark susceptibilities. Furthermore, for what concern the charm mass, we have found that even if a constant mass value $m_c=1.5$ $GeV$ gives a reasonable description of the charm quark susceptibility in a wide T range, still a fit to lQCD $\chi_c(T)$ shows that a decreasing $m_c(T)$ with values of  $m_c(T_c) \approx 2$ $GeV$ at $T_c$ [Case 3], is the best parametrization within a QPM approach.
Finally, $QPM_p$ with such a $m_c(T)$ entails at the same time a $D_s(T)$ that is quite close to the one from recent lQCD data, quite better than the one in the QPM standard modeling, especially in the non-perturbative region $T<2\, T_c$.

We have also presented a first calculation regarding the evolution of the charm quark distribution function in terms of D mesons $R_{AA}(p)$, for a system whose temperature decreases like hydro 1D.
In the extended model $QPM_p$ with the momentum dependent masses, we observe globally a slower dynamics wrt the simple $QPM$, with a smaller $R_{AA}$ in the low $p$ region. This could be significant considering that QPM in realistic simulations tend to overestimate $R_{AA}$ in the low $p_T$ region. This is an important aspect because lQCD explore the $D_s$ in the vanishing momentum limit and also upcoming experimental data, thanks to the new upgrades of both ALICE and CMS, will allow to access low $p_T$ observables with high precision. During the completion of the present work, we become aware that a new more recent evaluation of the $D_s(T)$ at finite charm and bottom mass has been worked out by combining lQCD to non-relativistic effective field theory approach \cite{Altenkort:2023eav}. The result shows a  weak mass dependence of $D_s(T)$: this is an aspect that should be further explored in the future comparing the mass dependence in the different phenomenological modeling. Also the quantitative validity of the non relativistic approximation especially for charm quark mass. A value $2\pi T D_s \simeq 1-2$ as found in \cite{Altenkort:2023oms} already at the charm quark mass seems quite small with respect to several phenomenological estimates. This is also an aspect we think should be focused and the momentum dependence of the transport coefficient may play a key role. In fact, it also should be considered that the estimate of $D_s$ from phenomenology till now comes mainly from comparing to experimental data at intermediate momentum and then extrapolating within the model at $p \rightarrow \, 0$. 

A next step in our approach will be to study the predictions in a realistic 3+1D simulation that include hadronization for both bottom and charm hadron to investigate if the $QPM_p$ supplies transport properties leading to a better agreement to experimental observables.

\subsection*{Acknowledgments}
V.G and S.P. acknowledge the funding from UniCT under ‘Linea di intervento 2’ (HQCDyn Grant) and PRIN2022 (Project code 2022SM5YAS) within Next Generation EU fundings and the support from the European Union's Horizon 2020 research. The authors acknowledge innovation program Strong 2020 under grant agreement No 824093.


\begin{thebibliography}{79}
\expandafter\ifx\csname natexlab\endcsname\relax\def\natexlab#1{#1}\fi
\expandafter\ifx\csname bibnamefont\endcsname\relax
  \def\bibnamefont#1{#1}\fi
\expandafter\ifx\csname bibfnamefont\endcsname\relax
  \def\bibfnamefont#1{#1}\fi
\expandafter\ifx\csname citenamefont\endcsname\relax
  \def\citenamefont#1{#1}\fi
\expandafter\ifx\csname url\endcsname\relax
  \def\url#1{\texttt{#1}}\fi
\expandafter\ifx\csname urlprefix\endcsname\relax\def\urlprefix{URL }\fi
\providecommand{\bibinfo}[2]{#2}
\providecommand{\eprint}[2][]{\url{#2}}

\bibitem[{\citenamefont{Borsanyi et~al.}(2014)\citenamefont{Borsanyi, Fodor, Hoelbling, Katz, Krieg, and Szabo}}]{Borsanyi:2013bia}
\bibinfo{author}{\bibfnamefont{S.}~\bibnamefont{Borsanyi}}, \bibinfo{author}{\bibfnamefont{Z.}~\bibnamefont{Fodor}}, \bibinfo{author}{\bibfnamefont{C.}~\bibnamefont{Hoelbling}}, \bibinfo{author}{\bibfnamefont{S.~D.} \bibnamefont{Katz}}, \bibinfo{author}{\bibfnamefont{S.}~\bibnamefont{Krieg}}, \bibnamefont{and} \bibinfo{author}{\bibfnamefont{K.~K.} \bibnamefont{Szabo}}, \bibinfo{journal}{Phys. Lett. B} \textbf{\bibinfo{volume}{730}}, \bibinfo{pages}{99} (\bibinfo{year}{2014}), \eprint{1309.5258}.

\bibitem[{\citenamefont{Soltz et~al.}(2015)\citenamefont{Soltz, DeTar, Karsch, Mukherjee, and Vranas}}]{Soltz:2015ula}
\bibinfo{author}{\bibfnamefont{R.~A.} \bibnamefont{Soltz}}, \bibinfo{author}{\bibfnamefont{C.}~\bibnamefont{DeTar}}, \bibinfo{author}{\bibfnamefont{F.}~\bibnamefont{Karsch}}, \bibinfo{author}{\bibfnamefont{S.}~\bibnamefont{Mukherjee}}, \bibnamefont{and} \bibinfo{author}{\bibfnamefont{P.}~\bibnamefont{Vranas}}, \bibinfo{journal}{Ann. Rev. Nucl. Part. Sci.} \textbf{\bibinfo{volume}{65}}, \bibinfo{pages}{379} (\bibinfo{year}{2015}), \eprint{1502.02296}.

\bibitem[{\citenamefont{Asakawa and Yazaki}(1989)}]{Asakawa:1989bq}
\bibinfo{author}{\bibfnamefont{M.}~\bibnamefont{Asakawa}} \bibnamefont{and} \bibinfo{author}{\bibfnamefont{K.}~\bibnamefont{Yazaki}}, \bibinfo{journal}{Nucl. Phys. A} \textbf{\bibinfo{volume}{504}}, \bibinfo{pages}{668} (\bibinfo{year}{1989}).

\bibitem[{\citenamefont{Fischer and Luecker}(2013)}]{Fischer:2012vc}
\bibinfo{author}{\bibfnamefont{C.~S.} \bibnamefont{Fischer}} \bibnamefont{and} \bibinfo{author}{\bibfnamefont{J.}~\bibnamefont{Luecker}}, \bibinfo{journal}{Phys. Lett. B} \textbf{\bibinfo{volume}{718}}, \bibinfo{pages}{1036} (\bibinfo{year}{2013}), \eprint{1206.5191}.

\bibitem[{\citenamefont{Herbst et~al.}(2011)\citenamefont{Herbst, Pawlowski, and Schaefer}}]{Herbst:2010rf}
\bibinfo{author}{\bibfnamefont{T.~K.} \bibnamefont{Herbst}}, \bibinfo{author}{\bibfnamefont{J.~M.} \bibnamefont{Pawlowski}}, \bibnamefont{and} \bibinfo{author}{\bibfnamefont{B.-J.} \bibnamefont{Schaefer}}, \bibinfo{journal}{Phys. Lett. B} \textbf{\bibinfo{volume}{696}}, \bibinfo{pages}{58} (\bibinfo{year}{2011}), \eprint{1008.0081}.

\bibitem[{\citenamefont{Schaefer et~al.}(2007)\citenamefont{Schaefer, Pawlowski, and Wambach}}]{Schaefer:2007pw}
\bibinfo{author}{\bibfnamefont{B.-J.} \bibnamefont{Schaefer}}, \bibinfo{author}{\bibfnamefont{J.~M.} \bibnamefont{Pawlowski}}, \bibnamefont{and} \bibinfo{author}{\bibfnamefont{J.}~\bibnamefont{Wambach}}, \bibinfo{journal}{Phys. Rev. D} \textbf{\bibinfo{volume}{76}}, \bibinfo{pages}{074023} (\bibinfo{year}{2007}), \eprint{0704.3234}.

\bibitem[{\citenamefont{Critelli et~al.}(2017)\citenamefont{Critelli, Noronha, Noronha-Hostler, Portillo, Ratti, and Rougemont}}]{Critelli:2017oub}
\bibinfo{author}{\bibfnamefont{R.}~\bibnamefont{Critelli}}, \bibinfo{author}{\bibfnamefont{J.}~\bibnamefont{Noronha}}, \bibinfo{author}{\bibfnamefont{J.}~\bibnamefont{Noronha-Hostler}}, \bibinfo{author}{\bibfnamefont{I.}~\bibnamefont{Portillo}}, \bibinfo{author}{\bibfnamefont{C.}~\bibnamefont{Ratti}}, \bibnamefont{and} \bibinfo{author}{\bibfnamefont{R.}~\bibnamefont{Rougemont}}, \bibinfo{journal}{Phys. Rev. D} \textbf{\bibinfo{volume}{96}}, \bibinfo{pages}{096026} (\bibinfo{year}{2017}), \eprint{1706.00455}.

\bibitem[{\citenamefont{Stephanov et~al.}(1998)\citenamefont{Stephanov, Rajagopal, and Shuryak}}]{Stephanov:1998dy}
\bibinfo{author}{\bibfnamefont{M.~A.} \bibnamefont{Stephanov}}, \bibinfo{author}{\bibfnamefont{K.}~\bibnamefont{Rajagopal}}, \bibnamefont{and} \bibinfo{author}{\bibfnamefont{E.~V.} \bibnamefont{Shuryak}}, \bibinfo{journal}{Phys. Rev. Lett.} \textbf{\bibinfo{volume}{81}}, \bibinfo{pages}{4816} (\bibinfo{year}{1998}), \eprint{hep-ph/9806219}.

\bibitem[{\citenamefont{Braaten and Pisarski}(1992)}]{Braaten:1991gm}
\bibinfo{author}{\bibfnamefont{E.}~\bibnamefont{Braaten}} \bibnamefont{and} \bibinfo{author}{\bibfnamefont{R.~D.} \bibnamefont{Pisarski}}, \bibinfo{journal}{Phys. Rev. D} \textbf{\bibinfo{volume}{45}}, \bibinfo{pages}{R1827} (\bibinfo{year}{1992}).

\bibitem[{\citenamefont{Blaizot and Iancu}(1994)}]{Blaizot:1993be}
\bibinfo{author}{\bibfnamefont{J.~P.} \bibnamefont{Blaizot}} \bibnamefont{and} \bibinfo{author}{\bibfnamefont{E.}~\bibnamefont{Iancu}}, \bibinfo{journal}{Nucl. Phys. B} \textbf{\bibinfo{volume}{417}}, \bibinfo{pages}{608} (\bibinfo{year}{1994}), \eprint{hep-ph/9306294}.

\bibitem[{\citenamefont{Andersen et~al.}(1999)\citenamefont{Andersen, Braaten, and Strickland}}]{Andersen:1999fw}
\bibinfo{author}{\bibfnamefont{J.~O.} \bibnamefont{Andersen}}, \bibinfo{author}{\bibfnamefont{E.}~\bibnamefont{Braaten}}, \bibnamefont{and} \bibinfo{author}{\bibfnamefont{M.}~\bibnamefont{Strickland}}, \bibinfo{journal}{Phys. Rev. Lett.} \textbf{\bibinfo{volume}{83}}, \bibinfo{pages}{2139} (\bibinfo{year}{1999}), \eprint{hep-ph/9902327}.

\bibitem[{\citenamefont{Andersen et~al.}(2002)\citenamefont{Andersen, Braaten, Petitgirard, and Strickland}}]{Andersen:2002ey}
\bibinfo{author}{\bibfnamefont{J.~O.} \bibnamefont{Andersen}}, \bibinfo{author}{\bibfnamefont{E.}~\bibnamefont{Braaten}}, \bibinfo{author}{\bibfnamefont{E.}~\bibnamefont{Petitgirard}}, \bibnamefont{and} \bibinfo{author}{\bibfnamefont{M.}~\bibnamefont{Strickland}}, \bibinfo{journal}{Phys. Rev. D} \textbf{\bibinfo{volume}{66}}, \bibinfo{pages}{085016} (\bibinfo{year}{2002}), \eprint{hep-ph/0205085}.

\bibitem[{\citenamefont{Blaizot et~al.}(1999)\citenamefont{Blaizot, Iancu, and Rebhan}}]{Blaizot:1999ap}
\bibinfo{author}{\bibfnamefont{J.~P.} \bibnamefont{Blaizot}}, \bibinfo{author}{\bibfnamefont{E.}~\bibnamefont{Iancu}}, \bibnamefont{and} \bibinfo{author}{\bibfnamefont{A.}~\bibnamefont{Rebhan}}, \bibinfo{journal}{Phys. Lett. B} \textbf{\bibinfo{volume}{470}}, \bibinfo{pages}{181} (\bibinfo{year}{1999}), \eprint{hep-ph/9910309}.

\bibitem[{\citenamefont{Andersen et~al.}(2011)\citenamefont{Andersen, Leganger, Strickland, and Su}}]{Andersen:2010wu}
\bibinfo{author}{\bibfnamefont{J.~O.} \bibnamefont{Andersen}}, \bibinfo{author}{\bibfnamefont{L.~E.} \bibnamefont{Leganger}}, \bibinfo{author}{\bibfnamefont{M.}~\bibnamefont{Strickland}}, \bibnamefont{and} \bibinfo{author}{\bibfnamefont{N.}~\bibnamefont{Su}}, \bibinfo{journal}{Phys. Lett. B} \textbf{\bibinfo{volume}{696}}, \bibinfo{pages}{468} (\bibinfo{year}{2011}), \eprint{1009.4644}.

\bibitem[{\citenamefont{Levai and Heinz}(1998)}]{Levai:1997yx}
\bibinfo{author}{\bibfnamefont{P.}~\bibnamefont{Levai}} \bibnamefont{and} \bibinfo{author}{\bibfnamefont{U.~W.} \bibnamefont{Heinz}}, \bibinfo{journal}{Phys. Rev. C} \textbf{\bibinfo{volume}{57}}, \bibinfo{pages}{1879} (\bibinfo{year}{1998}), \eprint{hep-ph/9710463}.

\bibitem[{\citenamefont{Peshier et~al.}(2002)\citenamefont{Peshier, Kampfer, and Soff}}]{Peshier:2002ww}
\bibinfo{author}{\bibfnamefont{A.}~\bibnamefont{Peshier}}, \bibinfo{author}{\bibfnamefont{B.}~\bibnamefont{Kampfer}}, \bibnamefont{and} \bibinfo{author}{\bibfnamefont{G.}~\bibnamefont{Soff}}, \bibinfo{journal}{Phys. Rev. D} \textbf{\bibinfo{volume}{66}}, \bibinfo{pages}{094003} (\bibinfo{year}{2002}), \eprint{hep-ph/0206229}.

\bibitem[{\citenamefont{Plumari et~al.}(2011)\citenamefont{Plumari, Alberico, Greco, and Ratti}}]{Plumari:2011mk}
\bibinfo{author}{\bibfnamefont{S.}~\bibnamefont{Plumari}}, \bibinfo{author}{\bibfnamefont{W.~M.} \bibnamefont{Alberico}}, \bibinfo{author}{\bibfnamefont{V.}~\bibnamefont{Greco}}, \bibnamefont{and} \bibinfo{author}{\bibfnamefont{C.}~\bibnamefont{Ratti}}, \bibinfo{journal}{Phys. Rev. D} \textbf{\bibinfo{volume}{84}}, \bibinfo{pages}{094004} (\bibinfo{year}{2011}).

\bibitem[{\citenamefont{Song et~al.}(2016)\citenamefont{Song, Berrehrah, Cabrera, Cassing, and Bratkovskaya}}]{Song:2015ykw}
\bibinfo{author}{\bibfnamefont{T.}~\bibnamefont{Song}}, \bibinfo{author}{\bibfnamefont{H.}~\bibnamefont{Berrehrah}}, \bibinfo{author}{\bibfnamefont{D.}~\bibnamefont{Cabrera}}, \bibinfo{author}{\bibfnamefont{W.}~\bibnamefont{Cassing}}, \bibnamefont{and} \bibinfo{author}{\bibfnamefont{E.}~\bibnamefont{Bratkovskaya}}, \bibinfo{journal}{Phys. Rev. C} \textbf{\bibinfo{volume}{93}}, \bibinfo{pages}{034906} (\bibinfo{year}{2016}), \eprint{1512.00891}.

\bibitem[{\citenamefont{Liu et~al.}(2024)\citenamefont{Liu, Wu, Cao, Qin, and Wang}}]{Liu:2023rfi}
\bibinfo{author}{\bibfnamefont{F.-L.} \bibnamefont{Liu}}, \bibinfo{author}{\bibfnamefont{X.-Y.} \bibnamefont{Wu}}, \bibinfo{author}{\bibfnamefont{S.}~\bibnamefont{Cao}}, \bibinfo{author}{\bibfnamefont{G.-Y.} \bibnamefont{Qin}}, \bibnamefont{and} \bibinfo{author}{\bibfnamefont{X.-N.} \bibnamefont{Wang}}, \bibinfo{journal}{Phys. Lett. B} \textbf{\bibinfo{volume}{848}}, \bibinfo{pages}{138355} (\bibinfo{year}{2024}), \eprint{2304.08787}.

\bibitem[{\citenamefont{Soloveva et~al.}(2023)\citenamefont{Soloveva, Palermo, and Bratkovskaya}}]{Soloveva:2023tvj}
\bibinfo{author}{\bibfnamefont{O.}~\bibnamefont{Soloveva}}, \bibinfo{author}{\bibfnamefont{A.}~\bibnamefont{Palermo}}, \bibnamefont{and} \bibinfo{author}{\bibfnamefont{E.}~\bibnamefont{Bratkovskaya}} (\bibinfo{year}{2023}), \eprint{2311.15984}.

\bibitem[{\citenamefont{Peshier and Cassing}(2005)}]{Peshier:2005pp}
\bibinfo{author}{\bibfnamefont{A.}~\bibnamefont{Peshier}} \bibnamefont{and} \bibinfo{author}{\bibfnamefont{W.}~\bibnamefont{Cassing}}, \bibinfo{journal}{Phys. Rev. Lett.} \textbf{\bibinfo{volume}{94}}, \bibinfo{pages}{172301} (\bibinfo{year}{2005}), \eprint{hep-ph/0502138}.

\bibitem[{\citenamefont{Cassing}(2007)}]{Cassing:2007nb}
\bibinfo{author}{\bibfnamefont{W.}~\bibnamefont{Cassing}}, \bibinfo{journal}{Nucl. Phys. A} \textbf{\bibinfo{volume}{795}}, \bibinfo{pages}{70} (\bibinfo{year}{2007}), \eprint{0707.3033}.

\bibitem[{\citenamefont{Berrehrah et~al.}(2014{\natexlab{a}})\citenamefont{Berrehrah, Bratkovskaya, Cassing, Gossiaux, Aichelin, and Bleicher}}]{Berrehrah:2013mua}
\bibinfo{author}{\bibfnamefont{H.}~\bibnamefont{Berrehrah}}, \bibinfo{author}{\bibfnamefont{E.}~\bibnamefont{Bratkovskaya}}, \bibinfo{author}{\bibfnamefont{W.}~\bibnamefont{Cassing}}, \bibinfo{author}{\bibfnamefont{P.~B.} \bibnamefont{Gossiaux}}, \bibinfo{author}{\bibfnamefont{J.}~\bibnamefont{Aichelin}}, \bibnamefont{and} \bibinfo{author}{\bibfnamefont{M.}~\bibnamefont{Bleicher}}, \bibinfo{journal}{Phys. Rev.} \textbf{\bibinfo{volume}{C89}}, \bibinfo{pages}{054901} (\bibinfo{year}{2014}{\natexlab{a}}).

\bibitem[{\citenamefont{Berrehrah et~al.}(2014{\natexlab{b}})\citenamefont{Berrehrah, Gossiaux, Aichelin, Cassing, and Bratkovskaya}}]{Berrehrah:2014kba}
\bibinfo{author}{\bibfnamefont{H.}~\bibnamefont{Berrehrah}}, \bibinfo{author}{\bibfnamefont{P.-B.} \bibnamefont{Gossiaux}}, \bibinfo{author}{\bibfnamefont{J.}~\bibnamefont{Aichelin}}, \bibinfo{author}{\bibfnamefont{W.}~\bibnamefont{Cassing}}, \bibnamefont{and} \bibinfo{author}{\bibfnamefont{E.}~\bibnamefont{Bratkovskaya}}, \bibinfo{journal}{Phys. Rev. C} \textbf{\bibinfo{volume}{90}}, \bibinfo{pages}{064906} (\bibinfo{year}{2014}{\natexlab{b}}), \eprint{1405.3243}.

\bibitem[{\citenamefont{Borsanyi et~al.}(2010)\citenamefont{Borsanyi, Endrodi, Fodor, Jakovac, Katz, Krieg, Ratti, and Szabo}}]{Borsanyi:2010cj}
\bibinfo{author}{\bibfnamefont{S.}~\bibnamefont{Borsanyi}}, \bibinfo{author}{\bibfnamefont{G.}~\bibnamefont{Endrodi}}, \bibinfo{author}{\bibfnamefont{Z.}~\bibnamefont{Fodor}}, \bibinfo{author}{\bibfnamefont{A.}~\bibnamefont{Jakovac}}, \bibinfo{author}{\bibfnamefont{S.~D.} \bibnamefont{Katz}}, \bibinfo{author}{\bibfnamefont{S.}~\bibnamefont{Krieg}}, \bibinfo{author}{\bibfnamefont{C.}~\bibnamefont{Ratti}}, \bibnamefont{and} \bibinfo{author}{\bibfnamefont{K.~K.} \bibnamefont{Szabo}}, \bibinfo{journal}{JHEP} \textbf{\bibinfo{volume}{11}}, \bibinfo{pages}{077} (\bibinfo{year}{2010}).

\bibitem[{\citenamefont{Borsanyi et~al.}(2016)}]{Borsanyi:2016ksw}
\bibinfo{author}{\bibfnamefont{S.}~\bibnamefont{Borsanyi}} \bibnamefont{et~al.}, \bibinfo{journal}{Nature} \textbf{\bibinfo{volume}{539}}, \bibinfo{pages}{69} (\bibinfo{year}{2016}), \eprint{1606.07494}.

\bibitem[{\citenamefont{Scardina et~al.}(2013)\citenamefont{Scardina, Colonna, Plumari, and Greco}}]{Scardina:2012mik}
\bibinfo{author}{\bibfnamefont{F.}~\bibnamefont{Scardina}}, \bibinfo{author}{\bibfnamefont{M.}~\bibnamefont{Colonna}}, \bibinfo{author}{\bibfnamefont{S.}~\bibnamefont{Plumari}}, \bibnamefont{and} \bibinfo{author}{\bibfnamefont{V.}~\bibnamefont{Greco}}, \bibinfo{journal}{Phys. Lett. B} \textbf{\bibinfo{volume}{724}}, \bibinfo{pages}{296} (\bibinfo{year}{2013}), \eprint{1202.2262}.

\bibitem[{\citenamefont{Ruggieri et~al.}(2015)\citenamefont{Ruggieri, Plumari, Scardina, and Greco}}]{Ruggieri:2015tsa}
\bibinfo{author}{\bibfnamefont{M.}~\bibnamefont{Ruggieri}}, \bibinfo{author}{\bibfnamefont{S.}~\bibnamefont{Plumari}}, \bibinfo{author}{\bibfnamefont{F.}~\bibnamefont{Scardina}}, \bibnamefont{and} \bibinfo{author}{\bibfnamefont{V.}~\bibnamefont{Greco}}, \bibinfo{journal}{Nucl. Phys. A} \textbf{\bibinfo{volume}{941}}, \bibinfo{pages}{201} (\bibinfo{year}{2015}), \eprint{1502.04596}.

\bibitem[{\citenamefont{Scardina et~al.}(2017)\citenamefont{Scardina, Das, Minissale, Plumari, and Greco}}]{Scardina:2017ipo}
\bibinfo{author}{\bibfnamefont{F.}~\bibnamefont{Scardina}}, \bibinfo{author}{\bibfnamefont{S.~K.} \bibnamefont{Das}}, \bibinfo{author}{\bibfnamefont{V.}~\bibnamefont{Minissale}}, \bibinfo{author}{\bibfnamefont{S.}~\bibnamefont{Plumari}}, \bibnamefont{and} \bibinfo{author}{\bibfnamefont{V.}~\bibnamefont{Greco}}, \bibinfo{journal}{Phys.\ Rev.\ C} \textbf{\bibinfo{volume}{96}}, \bibinfo{pages}{044905} (\bibinfo{year}{2017}), \eprint{1707.05452}.

\bibitem[{\citenamefont{Plumari}(2019)}]{Plumari:2019gwq}
\bibinfo{author}{\bibfnamefont{S.}~\bibnamefont{Plumari}}, \bibinfo{journal}{Eur. Phys. J. C} \textbf{\bibinfo{volume}{79}}, \bibinfo{pages}{2} (\bibinfo{year}{2019}).

\bibitem[{\citenamefont{Sambataro et~al.}(2022{\natexlab{a}})\citenamefont{Sambataro, Sun, Minissale, Plumari, and Greco}}]{Sambataro:2022sns}
\bibinfo{author}{\bibfnamefont{M.~L.} \bibnamefont{Sambataro}}, \bibinfo{author}{\bibfnamefont{Y.}~\bibnamefont{Sun}}, \bibinfo{author}{\bibfnamefont{V.}~\bibnamefont{Minissale}}, \bibinfo{author}{\bibfnamefont{S.}~\bibnamefont{Plumari}}, \bibnamefont{and} \bibinfo{author}{\bibfnamefont{V.}~\bibnamefont{Greco}}, \bibinfo{journal}{Eur. Phys. J. C} \textbf{\bibinfo{volume}{82}}, \bibinfo{pages}{833} (\bibinfo{year}{2022}{\natexlab{a}}), \eprint{2206.03160}.

\bibitem[{\citenamefont{Liu and Rapp}(2020)}]{Liu:2016ysz}
\bibinfo{author}{\bibfnamefont{S.~Y.~F.} \bibnamefont{Liu}} \bibnamefont{and} \bibinfo{author}{\bibfnamefont{R.}~\bibnamefont{Rapp}}, \bibinfo{journal}{Eur. Phys. J. A} \textbf{\bibinfo{volume}{56}}, \bibinfo{pages}{44} (\bibinfo{year}{2020}), \eprint{1612.09138}.

\bibitem[{\citenamefont{Liu and Rapp}(2018)}]{Liu:2017qah}
\bibinfo{author}{\bibfnamefont{S.~Y.~F.} \bibnamefont{Liu}} \bibnamefont{and} \bibinfo{author}{\bibfnamefont{R.}~\bibnamefont{Rapp}}, \bibinfo{journal}{Phys. Rev. C} \textbf{\bibinfo{volume}{97}}, \bibinfo{pages}{034918} (\bibinfo{year}{2018}), \eprint{1711.03282}.

\bibitem[{\citenamefont{Oliva et~al.}(2021)\citenamefont{Oliva, Plumari, and Greco}}]{Oliva:2020doe}
\bibinfo{author}{\bibfnamefont{L.}~\bibnamefont{Oliva}}, \bibinfo{author}{\bibfnamefont{S.}~\bibnamefont{Plumari}}, \bibnamefont{and} \bibinfo{author}{\bibfnamefont{V.}~\bibnamefont{Greco}}, \bibinfo{journal}{JHEP} \textbf{\bibinfo{volume}{05}}, \bibinfo{pages}{034} (\bibinfo{year}{2021}), \eprint{2009.11066}.

\bibitem[{\citenamefont{van Hees et~al.}(2006)\citenamefont{van Hees, Greco, and Rapp}}]{vanHees:2005wb}
\bibinfo{author}{\bibfnamefont{H.}~\bibnamefont{van Hees}}, \bibinfo{author}{\bibfnamefont{V.}~\bibnamefont{Greco}}, \bibnamefont{and} \bibinfo{author}{\bibfnamefont{R.}~\bibnamefont{Rapp}}, \bibinfo{journal}{Phys. Rev. C} \textbf{\bibinfo{volume}{73}}, \bibinfo{pages}{034913} (\bibinfo{year}{2006}), \eprint{nucl-th/0508055}.

\bibitem[{\citenamefont{van Hees et~al.}(2008)\citenamefont{van Hees, Mannarelli, Greco, and Rapp}}]{vanHees:2007me}
\bibinfo{author}{\bibfnamefont{H.}~\bibnamefont{van Hees}}, \bibinfo{author}{\bibfnamefont{M.}~\bibnamefont{Mannarelli}}, \bibinfo{author}{\bibfnamefont{V.}~\bibnamefont{Greco}}, \bibnamefont{and} \bibinfo{author}{\bibfnamefont{R.}~\bibnamefont{Rapp}}, \bibinfo{journal}{Phys. Rev. Lett.} \textbf{\bibinfo{volume}{100}}, \bibinfo{pages}{192301} (\bibinfo{year}{2008}), \eprint{0709.2884}.

\bibitem[{\citenamefont{Gossiaux and Aichelin}(2008)}]{Gossiaux:2008jv}
\bibinfo{author}{\bibfnamefont{P.~B.} \bibnamefont{Gossiaux}} \bibnamefont{and} \bibinfo{author}{\bibfnamefont{J.}~\bibnamefont{Aichelin}}, \bibinfo{journal}{Phys. Rev. C} \textbf{\bibinfo{volume}{78}}, \bibinfo{pages}{014904} (\bibinfo{year}{2008}), \eprint{0802.2525}.

\bibitem[{\citenamefont{Alberico et~al.}(2011)\citenamefont{Alberico, Beraudo, De~Pace, Molinari, Monteno, Nardi, and Prino}}]{Alberico:2011zy}
\bibinfo{author}{\bibfnamefont{W.~M.} \bibnamefont{Alberico}}, \bibinfo{author}{\bibfnamefont{A.}~\bibnamefont{Beraudo}}, \bibinfo{author}{\bibfnamefont{A.}~\bibnamefont{De~Pace}}, \bibinfo{author}{\bibfnamefont{A.}~\bibnamefont{Molinari}}, \bibinfo{author}{\bibfnamefont{M.}~\bibnamefont{Monteno}}, \bibinfo{author}{\bibfnamefont{M.}~\bibnamefont{Nardi}}, \bibnamefont{and} \bibinfo{author}{\bibfnamefont{F.}~\bibnamefont{Prino}}, \bibinfo{journal}{Eur. Phys. J. C} \textbf{\bibinfo{volume}{71}}, \bibinfo{pages}{1666} (\bibinfo{year}{2011}), \eprint{1101.6008}.

\bibitem[{\citenamefont{Uphoff et~al.}(2012)\citenamefont{Uphoff, Fochler, Xu, and Greiner}}]{Uphoff:2012gb}
\bibinfo{author}{\bibfnamefont{J.}~\bibnamefont{Uphoff}}, \bibinfo{author}{\bibfnamefont{O.}~\bibnamefont{Fochler}}, \bibinfo{author}{\bibfnamefont{Z.}~\bibnamefont{Xu}}, \bibnamefont{and} \bibinfo{author}{\bibfnamefont{C.}~\bibnamefont{Greiner}}, \bibinfo{journal}{Phys. Lett. B} \textbf{\bibinfo{volume}{717}}, \bibinfo{pages}{430} (\bibinfo{year}{2012}), \eprint{1205.4945}.

\bibitem[{\citenamefont{Lang et~al.}(2016)\citenamefont{Lang, van Hees, Steinheimer, Inghirami, and Bleicher}}]{Lang:2012nqy}
\bibinfo{author}{\bibfnamefont{T.}~\bibnamefont{Lang}}, \bibinfo{author}{\bibfnamefont{H.}~\bibnamefont{van Hees}}, \bibinfo{author}{\bibfnamefont{J.}~\bibnamefont{Steinheimer}}, \bibinfo{author}{\bibfnamefont{G.}~\bibnamefont{Inghirami}}, \bibnamefont{and} \bibinfo{author}{\bibfnamefont{M.}~\bibnamefont{Bleicher}}, \bibinfo{journal}{Phys. Rev. C} \textbf{\bibinfo{volume}{93}}, \bibinfo{pages}{014901} (\bibinfo{year}{2016}), \eprint{1211.6912}.

\bibitem[{\citenamefont{Song et~al.}(2015)\citenamefont{Song, Berrehrah, Cabrera, Torres-Rincon, Tolos, Cassing, and Bratkovskaya}}]{Song:2015sfa}
\bibinfo{author}{\bibfnamefont{T.}~\bibnamefont{Song}}, \bibinfo{author}{\bibfnamefont{H.}~\bibnamefont{Berrehrah}}, \bibinfo{author}{\bibfnamefont{D.}~\bibnamefont{Cabrera}}, \bibinfo{author}{\bibfnamefont{J.~M.} \bibnamefont{Torres-Rincon}}, \bibinfo{author}{\bibfnamefont{L.}~\bibnamefont{Tolos}}, \bibinfo{author}{\bibfnamefont{W.}~\bibnamefont{Cassing}}, \bibnamefont{and} \bibinfo{author}{\bibfnamefont{E.}~\bibnamefont{Bratkovskaya}}, \bibinfo{journal}{Phys. Rev. C} \textbf{\bibinfo{volume}{92}}, \bibinfo{pages}{014910} (\bibinfo{year}{2015}), \eprint{1503.03039}.

\bibitem[{\citenamefont{Das et~al.}(2014)\citenamefont{Das, Scardina, Plumari, and Greco}}]{Das:2013kea}
\bibinfo{author}{\bibfnamefont{S.~K.} \bibnamefont{Das}}, \bibinfo{author}{\bibfnamefont{F.}~\bibnamefont{Scardina}}, \bibinfo{author}{\bibfnamefont{S.}~\bibnamefont{Plumari}}, \bibnamefont{and} \bibinfo{author}{\bibfnamefont{V.}~\bibnamefont{Greco}}, \bibinfo{journal}{Phys. Rev. C} \textbf{\bibinfo{volume}{90}}, \bibinfo{pages}{044901} (\bibinfo{year}{2014}), \eprint{1312.6857}.

\bibitem[{\citenamefont{Cao et~al.}(2015)\citenamefont{Cao, Qin, and Bass}}]{Cao:2015hia}
\bibinfo{author}{\bibfnamefont{S.}~\bibnamefont{Cao}}, \bibinfo{author}{\bibfnamefont{G.-Y.} \bibnamefont{Qin}}, \bibnamefont{and} \bibinfo{author}{\bibfnamefont{S.~A.} \bibnamefont{Bass}}, \bibinfo{journal}{Phys. Rev. C} \textbf{\bibinfo{volume}{92}}, \bibinfo{pages}{024907} (\bibinfo{year}{2015}), \eprint{1505.01413}.

\bibitem[{\citenamefont{Das et~al.}(2015)\citenamefont{Das, Scardina, Plumari, and Greco}}]{Das:2015ana}
\bibinfo{author}{\bibfnamefont{S.~K.} \bibnamefont{Das}}, \bibinfo{author}{\bibfnamefont{F.}~\bibnamefont{Scardina}}, \bibinfo{author}{\bibfnamefont{S.}~\bibnamefont{Plumari}}, \bibnamefont{and} \bibinfo{author}{\bibfnamefont{V.}~\bibnamefont{Greco}}, \bibinfo{journal}{Phys. Lett. B} \textbf{\bibinfo{volume}{747}}, \bibinfo{pages}{260} (\bibinfo{year}{2015}), \eprint{1502.03757}.

\bibitem[{\citenamefont{Cao et~al.}(2018)\citenamefont{Cao, Luo, Qin, and Wang}}]{Cao:2017hhk}
\bibinfo{author}{\bibfnamefont{S.}~\bibnamefont{Cao}}, \bibinfo{author}{\bibfnamefont{T.}~\bibnamefont{Luo}}, \bibinfo{author}{\bibfnamefont{G.-Y.} \bibnamefont{Qin}}, \bibnamefont{and} \bibinfo{author}{\bibfnamefont{X.-N.} \bibnamefont{Wang}}, \bibinfo{journal}{Phys. Lett. B} \textbf{\bibinfo{volume}{777}}, \bibinfo{pages}{255} (\bibinfo{year}{2018}), \eprint{1703.00822}.

\bibitem[{\citenamefont{Das et~al.}(2017)\citenamefont{Das, Ruggieri, Scardina, Plumari, and Greco}}]{Das:2017dsh}
\bibinfo{author}{\bibfnamefont{S.~K.} \bibnamefont{Das}}, \bibinfo{author}{\bibfnamefont{M.}~\bibnamefont{Ruggieri}}, \bibinfo{author}{\bibfnamefont{F.}~\bibnamefont{Scardina}}, \bibinfo{author}{\bibfnamefont{S.}~\bibnamefont{Plumari}}, \bibnamefont{and} \bibinfo{author}{\bibfnamefont{V.}~\bibnamefont{Greco}}, \bibinfo{journal}{J. Phys. G} \textbf{\bibinfo{volume}{44}}, \bibinfo{pages}{095102} (\bibinfo{year}{2017}), \eprint{1701.05123}.

\bibitem[{\citenamefont{Sun et~al.}(2019)\citenamefont{Sun, Coci, Das, Plumari, Ruggieri, and Greco}}]{Sun:2019fud}
\bibinfo{author}{\bibfnamefont{Y.}~\bibnamefont{Sun}}, \bibinfo{author}{\bibfnamefont{G.}~\bibnamefont{Coci}}, \bibinfo{author}{\bibfnamefont{S.~K.} \bibnamefont{Das}}, \bibinfo{author}{\bibfnamefont{S.}~\bibnamefont{Plumari}}, \bibinfo{author}{\bibfnamefont{M.}~\bibnamefont{Ruggieri}}, \bibnamefont{and} \bibinfo{author}{\bibfnamefont{V.}~\bibnamefont{Greco}}, \bibinfo{journal}{Phys. Lett. B} \textbf{\bibinfo{volume}{798}}, \bibinfo{pages}{134933} (\bibinfo{year}{2019}), \eprint{1902.06254}.

\bibitem[{\citenamefont{Coci et~al.}(2019)\citenamefont{Coci, Oliva, Plumari, Das, and Greco}}]{Coci:2019nyr}
\bibinfo{author}{\bibfnamefont{G.}~\bibnamefont{Coci}}, \bibinfo{author}{\bibfnamefont{L.}~\bibnamefont{Oliva}}, \bibinfo{author}{\bibfnamefont{S.}~\bibnamefont{Plumari}}, \bibinfo{author}{\bibfnamefont{S.~K.} \bibnamefont{Das}}, \bibnamefont{and} \bibinfo{author}{\bibfnamefont{V.}~\bibnamefont{Greco}}, \bibinfo{journal}{Nucl. Phys. A} \textbf{\bibinfo{volume}{982}}, \bibinfo{pages}{189} (\bibinfo{year}{2019}), \eprint{1901.05394}.

\bibitem[{\citenamefont{Li et~al.}(2019)\citenamefont{Li, Wang, Wan, and Liao}}]{Li:2019wri}
\bibinfo{author}{\bibfnamefont{S.}~\bibnamefont{Li}}, \bibinfo{author}{\bibfnamefont{C.}~\bibnamefont{Wang}}, \bibinfo{author}{\bibfnamefont{R.}~\bibnamefont{Wan}}, \bibnamefont{and} \bibinfo{author}{\bibfnamefont{J.}~\bibnamefont{Liao}}, \bibinfo{journal}{Phys. Rev. C} \textbf{\bibinfo{volume}{99}}, \bibinfo{pages}{054909} (\bibinfo{year}{2019}), \eprint{1901.04600}.

\bibitem[{\citenamefont{Cao et~al.}(2019)}]{Cao:2018ews}
\bibinfo{author}{\bibfnamefont{S.}~\bibnamefont{Cao}} \bibnamefont{et~al.}, \bibinfo{journal}{Phys. Rev. C} \textbf{\bibinfo{volume}{99}}, \bibinfo{pages}{054907} (\bibinfo{year}{2019}), \eprint{1809.07894}.

\bibitem[{\citenamefont{Beraudo et~al.}(2018)}]{Rapp:2018qla}
\bibinfo{author}{\bibfnamefont{A.}~\bibnamefont{Beraudo}} \bibnamefont{et~al.}, \bibinfo{journal}{Nucl. Phys. A} \textbf{\bibinfo{volume}{979}}, \bibinfo{pages}{21} (\bibinfo{year}{2018}), \eprint{1803.03824}.

\bibitem[{\citenamefont{Ravagli and Rapp}(2007)}]{Ravagli:2007xx}
\bibinfo{author}{\bibfnamefont{L.}~\bibnamefont{Ravagli}} \bibnamefont{and} \bibinfo{author}{\bibfnamefont{R.}~\bibnamefont{Rapp}}, \bibinfo{journal}{Phys. Lett. B} \textbf{\bibinfo{volume}{655}}, \bibinfo{pages}{126} (\bibinfo{year}{2007}), \eprint{0705.0021}.

\bibitem[{\citenamefont{Sambataro et~al.}(2024)\citenamefont{Sambataro, Minissale, Plumari, and Greco}}]{Sambataro:2023tlv}
\bibinfo{author}{\bibfnamefont{M.~L.} \bibnamefont{Sambataro}}, \bibinfo{author}{\bibfnamefont{V.}~\bibnamefont{Minissale}}, \bibinfo{author}{\bibfnamefont{S.}~\bibnamefont{Plumari}}, \bibnamefont{and} \bibinfo{author}{\bibfnamefont{V.}~\bibnamefont{Greco}}, \bibinfo{journal}{Phys. Lett. B} \textbf{\bibinfo{volume}{849}}, \bibinfo{pages}{138480} (\bibinfo{year}{2024}), \eprint{2304.02953}.

\bibitem[{\citenamefont{Plumari et~al.}(2020)\citenamefont{Plumari, Coci, Minissale, Das, Sun, and Greco}}]{Plumari:2019hzp}
\bibinfo{author}{\bibfnamefont{S.}~\bibnamefont{Plumari}}, \bibinfo{author}{\bibfnamefont{G.}~\bibnamefont{Coci}}, \bibinfo{author}{\bibfnamefont{V.}~\bibnamefont{Minissale}}, \bibinfo{author}{\bibfnamefont{S.~K.} \bibnamefont{Das}}, \bibinfo{author}{\bibfnamefont{Y.}~\bibnamefont{Sun}}, \bibnamefont{and} \bibinfo{author}{\bibfnamefont{V.}~\bibnamefont{Greco}}, \bibinfo{journal}{Phys. Lett. B} \textbf{\bibinfo{volume}{805}}, \bibinfo{pages}{135460} (\bibinfo{year}{2020}), \eprint{1912.09350}.

\bibitem[{\citenamefont{Pooja et~al.}(2023)\citenamefont{Pooja, Das, Greco, and Ruggieri}}]{Pooja:2023gqt}
\bibinfo{author}{\bibnamefont{Pooja}}, \bibinfo{author}{\bibfnamefont{S.~K.} \bibnamefont{Das}}, \bibinfo{author}{\bibfnamefont{V.}~\bibnamefont{Greco}}, \bibnamefont{and} \bibinfo{author}{\bibfnamefont{M.}~\bibnamefont{Ruggieri}}, \bibinfo{journal}{Phys. Rev. D} \textbf{\bibinfo{volume}{108}}, \bibinfo{pages}{054026} (\bibinfo{year}{2023}), \eprint{2306.13749}.

\bibitem[{\citenamefont{Berrehrah et~al.}(2016{\natexlab{a}})\citenamefont{Berrehrah, Cassing, Bratkovskaya, and Steinert}}]{Berrehrah:2015vhe}
\bibinfo{author}{\bibfnamefont{H.}~\bibnamefont{Berrehrah}}, \bibinfo{author}{\bibfnamefont{W.}~\bibnamefont{Cassing}}, \bibinfo{author}{\bibfnamefont{E.}~\bibnamefont{Bratkovskaya}}, \bibnamefont{and} \bibinfo{author}{\bibfnamefont{T.}~\bibnamefont{Steinert}}, \bibinfo{journal}{Phys. Rev. C} \textbf{\bibinfo{volume}{93}}, \bibinfo{pages}{044914} (\bibinfo{year}{2016}{\natexlab{a}}), \eprint{1512.06909}.

\bibitem[{\citenamefont{Berrehrah et~al.}(2016{\natexlab{b}})\citenamefont{Berrehrah, Bratkovskaya, Steinert, and Cassing}}]{Berrehrah:2016vzw}
\bibinfo{author}{\bibfnamefont{H.}~\bibnamefont{Berrehrah}}, \bibinfo{author}{\bibfnamefont{E.}~\bibnamefont{Bratkovskaya}}, \bibinfo{author}{\bibfnamefont{T.}~\bibnamefont{Steinert}}, \bibnamefont{and} \bibinfo{author}{\bibfnamefont{W.}~\bibnamefont{Cassing}}, \bibinfo{journal}{Int. J. Mod. Phys. E} \textbf{\bibinfo{volume}{25}}, \bibinfo{pages}{1642003} (\bibinfo{year}{2016}{\natexlab{b}}), \eprint{1605.02371}.

\bibitem[{\citenamefont{Peshier et~al.}(1996)\citenamefont{Peshier, Kampfer, Pavlenko, and Soff}}]{Peshier:1995ty}
\bibinfo{author}{\bibfnamefont{A.}~\bibnamefont{Peshier}}, \bibinfo{author}{\bibfnamefont{B.}~\bibnamefont{Kampfer}}, \bibinfo{author}{\bibfnamefont{O.~P.} \bibnamefont{Pavlenko}}, \bibnamefont{and} \bibinfo{author}{\bibfnamefont{G.}~\bibnamefont{Soff}}, \bibinfo{journal}{Phys. Rev. D} \textbf{\bibinfo{volume}{54}}, \bibinfo{pages}{2399} (\bibinfo{year}{1996}).

\bibitem[{\citenamefont{Bazavov et~al.}(2014)}]{Bazavov:2014yba}
\bibinfo{author}{\bibfnamefont{A.}~\bibnamefont{Bazavov}} \bibnamefont{et~al.}, \bibinfo{journal}{Phys.\ Lett.\ B} \textbf{\bibinfo{volume}{737}}, \bibinfo{pages}{210} (\bibinfo{year}{2014}), \eprint{1404.4043}.

\bibitem[{\citenamefont{Petreczky et~al.}(2009)\citenamefont{Petreczky, Hegde, and Velytsky}}]{Petreczky:2009cr}
\bibinfo{author}{\bibfnamefont{P.}~\bibnamefont{Petreczky}}, \bibinfo{author}{\bibfnamefont{P.}~\bibnamefont{Hegde}}, \bibnamefont{and} \bibinfo{author}{\bibfnamefont{A.}~\bibnamefont{Velytsky}} (\bibinfo{collaboration}{RBC-Bielefeld}), \bibinfo{journal}{PoS} \textbf{\bibinfo{volume}{LAT2009}}, \bibinfo{pages}{159} (\bibinfo{year}{2009}), \eprint{0911.0196}.

\bibitem[{\citenamefont{Bellwied et~al.}(2015)\citenamefont{Bellwied, Borsanyi, Fodor, Katz, Pasztor, Ratti, and Szabo}}]{Bellwied:2015lba}
\bibinfo{author}{\bibfnamefont{R.}~\bibnamefont{Bellwied}}, \bibinfo{author}{\bibfnamefont{S.}~\bibnamefont{Borsanyi}}, \bibinfo{author}{\bibfnamefont{Z.}~\bibnamefont{Fodor}}, \bibinfo{author}{\bibfnamefont{S.~D.} \bibnamefont{Katz}}, \bibinfo{author}{\bibfnamefont{A.}~\bibnamefont{Pasztor}}, \bibinfo{author}{\bibfnamefont{C.}~\bibnamefont{Ratti}}, \bibnamefont{and} \bibinfo{author}{\bibfnamefont{K.~K.} \bibnamefont{Szabo}}, \bibinfo{journal}{Phys. Rev. D} \textbf{\bibinfo{volume}{92}}, \bibinfo{pages}{114505} (\bibinfo{year}{2015}), \eprint{1507.04627}.

\bibitem[{\citenamefont{Tang et~al.}(2023)\citenamefont{Tang, Mukherjee, Petreczky, and Rapp}}]{ZhanduoTang:2023ewm}
\bibinfo{author}{\bibfnamefont{Z.}~\bibnamefont{Tang}}, \bibinfo{author}{\bibfnamefont{S.}~\bibnamefont{Mukherjee}}, \bibinfo{author}{\bibfnamefont{P.}~\bibnamefont{Petreczky}}, \bibnamefont{and} \bibinfo{author}{\bibfnamefont{R.}~\bibnamefont{Rapp}} (\bibinfo{year}{2023}), \eprint{2310.18864}.

\bibitem[{\citenamefont{Fischer and Alkofer}(2003)}]{Fischer:2003rp}
\bibinfo{author}{\bibfnamefont{C.~S.} \bibnamefont{Fischer}} \bibnamefont{and} \bibinfo{author}{\bibfnamefont{R.}~\bibnamefont{Alkofer}}, \bibinfo{journal}{Phys. Rev. D} \textbf{\bibinfo{volume}{67}}, \bibinfo{pages}{094020} (\bibinfo{year}{2003}), \eprint{hep-ph/0301094}.

\bibitem[{\citenamefont{Fischer}(2006)}]{Fischer:2006ub}
\bibinfo{author}{\bibfnamefont{C.~S.} \bibnamefont{Fischer}}, \bibinfo{journal}{J. Phys. G} \textbf{\bibinfo{volume}{32}}, \bibinfo{pages}{R253} (\bibinfo{year}{2006}), \eprint{hep-ph/0605173}.

\bibitem[{\citenamefont{Fischer et~al.}(2014)\citenamefont{Fischer, Luecker, and Welzbacher}}]{Fischer:2014ata}
\bibinfo{author}{\bibfnamefont{C.~S.} \bibnamefont{Fischer}}, \bibinfo{author}{\bibfnamefont{J.}~\bibnamefont{Luecker}}, \bibnamefont{and} \bibinfo{author}{\bibfnamefont{C.~A.} \bibnamefont{Welzbacher}}, \bibinfo{journal}{Phys. Rev. D} \textbf{\bibinfo{volume}{90}}, \bibinfo{pages}{034022} (\bibinfo{year}{2014}), \eprint{1405.4762}.

\bibitem[{\citenamefont{Mueller et~al.}(2010)\citenamefont{Mueller, Fischer, and Nickel}}]{Mueller:2010ah}
\bibinfo{author}{\bibfnamefont{J.~A.} \bibnamefont{Mueller}}, \bibinfo{author}{\bibfnamefont{C.~S.} \bibnamefont{Fischer}}, \bibnamefont{and} \bibinfo{author}{\bibfnamefont{D.}~\bibnamefont{Nickel}}, \bibinfo{journal}{Eur. Phys. J. C} \textbf{\bibinfo{volume}{70}}, \bibinfo{pages}{1037} (\bibinfo{year}{2010}), \eprint{1009.3762}.

\bibitem[{\citenamefont{Bazavov et~al.}(2009)}]{Bazavov:2009zn}
\bibinfo{author}{\bibfnamefont{A.}~\bibnamefont{Bazavov}} \bibnamefont{et~al.}, \bibinfo{journal}{Phys. Rev. D} \textbf{\bibinfo{volume}{80}}, \bibinfo{pages}{014504} (\bibinfo{year}{2009}), \eprint{0903.4379}.

\bibitem[{\citenamefont{Svetitsky}(1988)}]{Svetitsky:1987gq}
\bibinfo{author}{\bibfnamefont{B.}~\bibnamefont{Svetitsky}}, \bibinfo{journal}{Phys. Rev. D} \textbf{\bibinfo{volume}{37}}, \bibinfo{pages}{2484} (\bibinfo{year}{1988}).

\bibitem[{\citenamefont{Sambataro et~al.}(2020)\citenamefont{Sambataro, Plumari, and Greco}}]{Sambataro:2020pge}
\bibinfo{author}{\bibfnamefont{M.~L.} \bibnamefont{Sambataro}}, \bibinfo{author}{\bibfnamefont{S.}~\bibnamefont{Plumari}}, \bibnamefont{and} \bibinfo{author}{\bibfnamefont{V.}~\bibnamefont{Greco}}, \bibinfo{journal}{Eur. Phys. J. C} \textbf{\bibinfo{volume}{80}}, \bibinfo{pages}{1140} (\bibinfo{year}{2020}), \eprint{2005.14470}.

\bibitem[{\citenamefont{Banerjee et~al.}(2012)\citenamefont{Banerjee, Datta, Gavai, and Majumdar}}]{Banerjee:2011ra}
\bibinfo{author}{\bibfnamefont{D.}~\bibnamefont{Banerjee}}, \bibinfo{author}{\bibfnamefont{S.}~\bibnamefont{Datta}}, \bibinfo{author}{\bibfnamefont{R.}~\bibnamefont{Gavai}}, \bibnamefont{and} \bibinfo{author}{\bibfnamefont{P.}~\bibnamefont{Majumdar}}, \bibinfo{journal}{Phys. Rev. D} \textbf{\bibinfo{volume}{85}}, \bibinfo{pages}{014510} (\bibinfo{year}{2012}), \eprint{1109.5738}.

\bibitem[{\citenamefont{Kaczmarek}(2014)}]{Kaczmarek:2014jga}
\bibinfo{author}{\bibfnamefont{O.}~\bibnamefont{Kaczmarek}}, \bibinfo{journal}{Nucl. Phys. A} \textbf{\bibinfo{volume}{931}}, \bibinfo{pages}{633} (\bibinfo{year}{2014}), \eprint{1409.3724}.

\bibitem[{\citenamefont{Francis et~al.}(2015)\citenamefont{Francis, Kaczmarek, Laine, Neuhaus, and Ohno}}]{Francis:2015daa}
\bibinfo{author}{\bibfnamefont{A.}~\bibnamefont{Francis}}, \bibinfo{author}{\bibfnamefont{O.}~\bibnamefont{Kaczmarek}}, \bibinfo{author}{\bibfnamefont{M.}~\bibnamefont{Laine}}, \bibinfo{author}{\bibfnamefont{T.}~\bibnamefont{Neuhaus}}, \bibnamefont{and} \bibinfo{author}{\bibfnamefont{H.}~\bibnamefont{Ohno}}, \bibinfo{journal}{Phys. Rev. D} \textbf{\bibinfo{volume}{92}}, \bibinfo{pages}{116003} (\bibinfo{year}{2015}), \eprint{1508.04543}.

\bibitem[{\citenamefont{Brambilla et~al.}(2020)\citenamefont{Brambilla, Leino, Petreczky, and Vairo}}]{Brambilla:2020siz}
\bibinfo{author}{\bibfnamefont{N.}~\bibnamefont{Brambilla}}, \bibinfo{author}{\bibfnamefont{V.}~\bibnamefont{Leino}}, \bibinfo{author}{\bibfnamefont{P.}~\bibnamefont{Petreczky}}, \bibnamefont{and} \bibinfo{author}{\bibfnamefont{A.}~\bibnamefont{Vairo}}, \bibinfo{journal}{Phys. Rev. D} \textbf{\bibinfo{volume}{102}}, \bibinfo{pages}{074503} (\bibinfo{year}{2020}), \eprint{2007.10078}.

\bibitem[{\citenamefont{Altenkort et~al.}(2023)\citenamefont{Altenkort, Kaczmarek, Larsen, Mukherjee, Petreczky, Shu, and Stendebach}}]{Altenkort:2023oms}
\bibinfo{author}{\bibfnamefont{L.}~\bibnamefont{Altenkort}}, \bibinfo{author}{\bibfnamefont{O.}~\bibnamefont{Kaczmarek}}, \bibinfo{author}{\bibfnamefont{R.}~\bibnamefont{Larsen}}, \bibinfo{author}{\bibfnamefont{S.}~\bibnamefont{Mukherjee}}, \bibinfo{author}{\bibfnamefont{P.}~\bibnamefont{Petreczky}}, \bibinfo{author}{\bibfnamefont{H.-T.} \bibnamefont{Shu}}, \bibnamefont{and} \bibinfo{author}{\bibfnamefont{S.}~\bibnamefont{Stendebach}} (\bibinfo{year}{2023}), \eprint{2302.08501}.

\bibitem[{\citenamefont{Sambataro et~al.}(2022{\natexlab{b}})\citenamefont{Sambataro, Plumari, Sun, Minissale, and Greco}}]{Sambataro:2022xzx}
\bibinfo{author}{\bibfnamefont{M.~L.} \bibnamefont{Sambataro}}, \bibinfo{author}{\bibfnamefont{S.}~\bibnamefont{Plumari}}, \bibinfo{author}{\bibfnamefont{Y.}~\bibnamefont{Sun}}, \bibinfo{author}{\bibfnamefont{V.}~\bibnamefont{Minissale}}, \bibnamefont{and} \bibinfo{author}{\bibfnamefont{V.}~\bibnamefont{Greco}}, \bibinfo{journal}{EPJ Web Conf.} \textbf{\bibinfo{volume}{259}}, \bibinfo{pages}{10016} (\bibinfo{year}{2022}{\natexlab{b}}).

\bibitem[{\citenamefont{Plumari et~al.}(2018)\citenamefont{Plumari, Minissale, Das, Coci, and Greco}}]{Plumari:2017ntm}
\bibinfo{author}{\bibfnamefont{S.}~\bibnamefont{Plumari}}, \bibinfo{author}{\bibfnamefont{V.}~\bibnamefont{Minissale}}, \bibinfo{author}{\bibfnamefont{S.~K.} \bibnamefont{Das}}, \bibinfo{author}{\bibfnamefont{G.}~\bibnamefont{Coci}}, \bibnamefont{and} \bibinfo{author}{\bibfnamefont{V.}~\bibnamefont{Greco}}, \bibinfo{journal}{Eur. Phys. J. C} \textbf{\bibinfo{volume}{78}}, \bibinfo{pages}{348} (\bibinfo{year}{2018}), \eprint{1712.00730}.

\bibitem[{\citenamefont{Minissale et~al.}(2021)\citenamefont{Minissale, Plumari, and Greco}}]{Minissale:2020bif}
\bibinfo{author}{\bibfnamefont{V.}~\bibnamefont{Minissale}}, \bibinfo{author}{\bibfnamefont{S.}~\bibnamefont{Plumari}}, \bibnamefont{and} \bibinfo{author}{\bibfnamefont{V.}~\bibnamefont{Greco}}, \bibinfo{journal}{Phys. Lett. B} \textbf{\bibinfo{volume}{821}}, \bibinfo{pages}{136622} (\bibinfo{year}{2021}), \eprint{2012.12001}.

\bibitem[{\citenamefont{Minissale et~al.}(2024)\citenamefont{Minissale, Plumari, Sun, and Greco}}]{Minissale:2023dct}
\bibinfo{author}{\bibfnamefont{V.}~\bibnamefont{Minissale}}, \bibinfo{author}{\bibfnamefont{S.}~\bibnamefont{Plumari}}, \bibinfo{author}{\bibfnamefont{Y.}~\bibnamefont{Sun}}, \bibnamefont{and} \bibinfo{author}{\bibfnamefont{V.}~\bibnamefont{Greco}}, \bibinfo{journal}{Eur. Phys. J. C} \textbf{\bibinfo{volume}{84}}, \bibinfo{pages}{228} (\bibinfo{year}{2024}), \eprint{2305.03687}.

\bibitem[{\citenamefont{Altenkort et~al.}(2024)\citenamefont{Altenkort, de~la Cruz, Kaczmarek, Larsen, Moore, Mukherjee, Petreczky, Shu, and Stendebach}}]{Altenkort:2023eav}
\bibinfo{author}{\bibfnamefont{L.}~\bibnamefont{Altenkort}}, \bibinfo{author}{\bibfnamefont{D.}~\bibnamefont{de~la Cruz}}, \bibinfo{author}{\bibfnamefont{O.}~\bibnamefont{Kaczmarek}}, \bibinfo{author}{\bibfnamefont{R.}~\bibnamefont{Larsen}}, \bibinfo{author}{\bibfnamefont{G.~D.} \bibnamefont{Moore}}, \bibinfo{author}{\bibfnamefont{S.}~\bibnamefont{Mukherjee}}, \bibinfo{author}{\bibfnamefont{P.}~\bibnamefont{Petreczky}}, \bibinfo{author}{\bibfnamefont{H.-T.} \bibnamefont{Shu}}, \bibnamefont{and} \bibinfo{author}{\bibfnamefont{S.}~\bibnamefont{Stendebach}} (\bibinfo{collaboration}{HotQCD}), \bibinfo{journal}{Phys. Rev. Lett.} \textbf{\bibinfo{volume}{132}}, \bibinfo{pages}{051902} (\bibinfo{year}{2024}), \eprint{2311.01525}.

\end{thebibliography}

\end{document}